\pdfoutput=1
\documentclass[prl,twocolumn,superscriptaddress]{revtex4-2}
\usepackage{graphicx,empheq,amssymb,scalerel,tikz}
\renewcommand{\section}[1]{{\par\it #1.---}\ignorespaces}
\usepackage[colorlinks=true, citecolor=blue, urlcolor=blue, linkcolor=blue ]{hyperref}
\usetikzlibrary{svg.path}
\definecolor{orcidlogocol}{HTML}{A6CE39}
\tikzset{
	orcidlogo/.pic={
		\fill[orcidlogocol] svg{M256,128c0,70.7-57.3,128-128,128C57.3,256,0,198.7,0,128C0,57.3,57.3,0,128,0C198.7,0,256,57.3,256,128z};
		\fill[white] svg{M86.3,186.2H70.9V79.1h15.4v48.4V186.2z}
		svg{M108.9,79.1h41.6c39.6,0,57,28.3,57,53.6c0,27.5-21.5,53.6-56.8,53.6h-41.8V79.1z M124.3,172.4h24.5c34.9,0,42.9-26.5,42.9-39.7c0-21.5-13.7-39.7-43.7-39.7h-23.7V172.4z}
		svg{M88.7,56.8c0,5.5-4.5,10.1-10.1,10.1c-5.6,0-10.1-4.6-10.1-10.1c0-5.6,4.5-10.1,10.1-10.1C84.2,46.7,88.7,51.3,88.7,56.8z};}}
\newcommand\orcid[1]{\href{https://orcid.org/#1}{\mbox{\scalerel*{\begin{tikzpicture}[yscale=-1,transform shape]\pic{orcidlogo};\end{tikzpicture}}{|}}}}

\begin{document}
\title{Non-Hermitian Weyl semimetal and its Floquet engineering}
\author{Hong Wu\orcid{0000-0003-3276-7823}}
\affiliation{Lanzhou Center for Theoretical Physics, Key Laboratory of Theoretical Physics of Gansu Province, Lanzhou University, Lanzhou 730000, China}
\author{Jun-Hong An\orcid{0000-0002-3475-0729}}\email{anjhong@lzu.edu.cn}
\affiliation{Lanzhou Center for Theoretical Physics, Key Laboratory of Theoretical Physics of Gansu Province, Lanzhou University, Lanzhou 730000, China}
\begin{abstract}
It is generally believed that non-Hermiticity can transform Weyl semimetals into Weyl-exceptional-ring semimetals. However, this belief is from the systems without skin effect. We investigate the non-Hermitian Weyl semimental and its Floquet engineering in a system with skin effect, which breaks the bulk-boundary correspondence in its Hermitian counterpart. It is found in both the static and periodically driven cases that the skin effect makes this general belief no longer valid. We discover that exotic non-Hermitian topological matters, e.g., a composite phase of Weyl semimetal and topological insulator with the coexisting Fermi arc and chiral boundary states, a widely tunable Hall conductivity with multiple quantized plateaus, and a Weyl semimetal with anomalous Fermi arcs formed by the crossing of gapped bound state, can be generated by the Floquet engineering. Revealing the leading role of the skin effect in determining the feature of a semimental, our result supplies a useful way to artificially synthesize exotic non-Hermitian Weyl semimetals by periodic driving.
\end{abstract}
\maketitle

\section{Introduction}
Non-Hermitian topological phases have attracted much attention \cite{PhysRevLett.116.133903,PhysRevA.97.052115,PhysRevLett.121.086803,PhysRevLett.121.026808,PhysRevLett.123.066404,PhysRevB.99.081103,PhysRevX.9.041015,PhysRevLett.124.056802,PhysRevLett.125.126402,PhysRevLett.125.186802,PhysRevLett.118.040401,PhysRevLett.120.146402,PhysRevLett.125.033603,PhysRevLett.125.226402,PhysRevLett.122.237601,PhysRevLett.125.206402,PhysRevLett.115.040402,PhysRevX.6.021007,PhysRevLett.121.124501,Xiao2020,PhysRevA.99.062122,Helbig2020,Weidemann2020,PhysRevLett.127.090501}. People desires to know if the well-developed topological phases in Hermitian systems can be generalized to non-Hermitian cases. This is particularly nontrivial due to the non-Hermitian skin effect induced breakdown of bulk-boundary correspondence (BBC). Non-Hermiticity also opens a novel window in applying topological phases to design lasers \cite{Hodaei975,Feng972,Hararieaar4003}, invisible media \cite{Regensburger2012}, and sensing \cite{EP2017,Chen2017}. A recent effort has been underway to generalize Hermitian topological semimetals \cite{RevModPhys.90.015001,PhysRevB.84.235126,PhysRevLett.125.146401,Xu2015,PhysRevX.5.031013,PhysRevLett.120.146602,PhysRevLett.121.106403,PhysRevLett.120.026402,PhysRevB.96.041102,PhysRevB.97.155140,PhysRevLett.116.127202,PhysRevB.101.195412,PhysRevLett.119.196403} to non-Hermitian systems \cite{PhysRevB.97.041203,PhysRevB.99.041406,PhysRevB.99.075130,PhysRevB.100.245116,PhysRevB.102.045412,PhysRevLett.120.146601,PhysRevLett.124.236403,PhysRevLett.126.010401,PhysRevB.104.L161117,PhysRevResearch.3.013015,PhysRevLett.123.066405,PhysRevLett.124.073603}. It was revealed that the non-Hermiticity can transform a Weyl semimetal into a Weyl-exceptional-ring semimetal \cite{PhysRevB.97.075128,Cerjan2019,PhysRevB.100.245205,PhysRevResearch.2.043268,PhysRevLett.127.196801,PhysRevA.103.L051101}.
Various non-Hermitian semimetals in different structures, including exceptional links \cite{PhysRevB.99.081102} and knots \cite{PhysRevB.99.161115,PhysRevLett.124.186402}, were found. However, these systems have no skin effect. The skin effect generally causes the mismatching of the exceptional points (EPs) under the open boundary condition (OBC) and periodic boundary condition (PBC), which invalidates the topological characterization by the Bloch-band theory. Therefore, a natural question is whether the non-Hermiticity induced inter-conversion of Weyl EPs and rings is still true in non-Hermitian systems with the skin effect.

Coherent control via periodic driving dubbed as Floquet engineering has become a versatile tool in artificially creating novel topological phases in systems of ultracold atoms \cite{RevModPhys.89.011004,PhysRevLett.116.205301}, photonics \cite{Rechtsman2013,PhysRevLett.122.173901}, circuit QED systems \cite{Roushan2017,PhysRevLett.119.093901}, and graphene \cite{McIver2020}. Many exotic phases absent in static systems have been synthesized by Floquet engineering \cite{PhysRevLett.117.087402,PhysRevA.102.062201,PhysRevB.102.041119,PhysRevA.100.053608,PhysRevB.100.045423,PhysRevLett.123.190403,PhysRevB.101.045415,PhysRevB.103.L041115,PhysRevB.96.041126,PhysRevLett.123.190403}. The key role played by
periodic driving is changing symmetry and inducing an effective long-range hopping in lattice systems \cite{PhysRevB.87.201109,PhysRevB.93.184306,PhysRevA.100.023622,PhysRevLett.121.036401}, which efficiently decreases the difficulty in fabricating specific interactions in natural materials. An interesting question becomes: How can the non-Hermitian topological semimetals benefit from periodic driving? Although Floquet engineering to Weyl-exceptional-ring semimetals have been studied, those systems did not have the skin effect \cite{PhysRevB.102.205423,PhysRevA.102.062201}. Thus, a general study on Floquet engineering to  non-Hermitian topological semimetals is still lacking.

Here, we investigate the non-Hermitian Weyl semimetal (NHWS) and its Floquet engineering in a system with the skin effect. It is interesting to find that the skin effect invalidates the general belief that the non-Hermiticity could convert the Weyl points into exceptional rings. Via Floquet engineering, we discover an exotic composite topological matters of NHWS and topological insulator with coexisting surface Fermi arc and chiral boundary states and with an enhanced Hall conductivity with multiple quantized plateaus compared to the static case. We also find an exotic NHWS with anomalous Fermi arc formed by the crossings of the gapped bound states instead of the generally believed gapless chiral boundary states. Our results reveal the distinguished role of the skin effect, Floquet engineering, and their interplay in determining the feature of the NHWSs as well as offer us a useful way to artificially create exotic NHWSs absent in natural materials.

\section{Static NHWS}
We consider a NHWS model on a cubic lattice. Its Hamiltonian is
\begin{eqnarray}
\hat{H}&=&\sum_{\bf n}\Big\{{1\over 2}\Big[{f_x}\hat{C}^{\dag}_{{\bf n}+\bf{x}}(\tau_x-i{\tau_y})\hat{C}_{{\bf n}}+{f_y}\hat{C}^{\dag}_{{\bf n}+\bf{y}}(\tau_x-i{\tau_z})\hat{C}_{{\bf n}}\nonumber\\
&&+{f_z}\hat{C}^{\dag}_{{\bf n}+\bf{z}}\tau_x\hat{C}_{{\bf n}}+\text{H.c.}\Big]+\hat{C}^{\dag}_{{\bf n}}(m\tau_x+i\frac{\gamma}{2}\tau_y)\hat{C}_{{\bf n}}\Big\},\label{Haml}
\end{eqnarray}
where $\hat{C}_{\bf n}=(\hat{c}_{{\bf n},A},\hat{c}_{{\bf n},B})^T$; $\hat{c}_{{\bf n},j}$ is the annihilation operator at $j=A,B$ sublattice of the site ${\bf n}=(n_x,n_y,n_z)$; $f_\alpha$ are the hopping rates along the $\alpha=x$, $y$, and $z$ directions with the unit vectors $\bf{x}$, $\bf{y}$, and $\bf{z}$; $\tau_\alpha$ are the Pauli matrices acting on the sublattice degrees of freedom; $m$ is the mass term; and the non-Hermitian term $\gamma$ is the nonreciprocal intracell hopping rate. Under the PBC, Eq. \eqref{Haml} reads $\hat{H}=\sum_{\bf k}{\hat{\bf C}}^\dag_{\bf k}\mathcal{H}({\bf k}){\hat{\bf C}}_{\bf k}$ with $\hat{\bf C}_{\bf k}=(\hat{\tilde c}_{{\bf k},A},\hat{\tilde{c}}_{{\bf k},B})^T$ and
\begin{eqnarray}
\mathcal{H}(\mathbf{k})&=&(f_x\cos k_x+F_{k_y,k_z})\sigma_x\nonumber\\
&&+(f_x\sin k_x+i\frac{\gamma}{2})\sigma_y+f_y\sin k_y\sigma_z,~~~~\label{Hamt}
\end{eqnarray}
where $\sigma_\alpha$ are the Pauli matrices and $F_{k_y,k_z}=m+f_y\cos k_y+f_z\cos k_z$. A related model in the Hermitian case was studied in Ref. \cite{PhysRevA.92.013632}. The unique feature of the non-Hermitian system is the skin effect \cite{PhysRevLett.121.086803}, which breaks the BBC manifesting in the mismatching of the EPs under the OBC and PBC.

\begin{figure}[tbp]
\centering
\includegraphics[width=0.97\columnwidth]{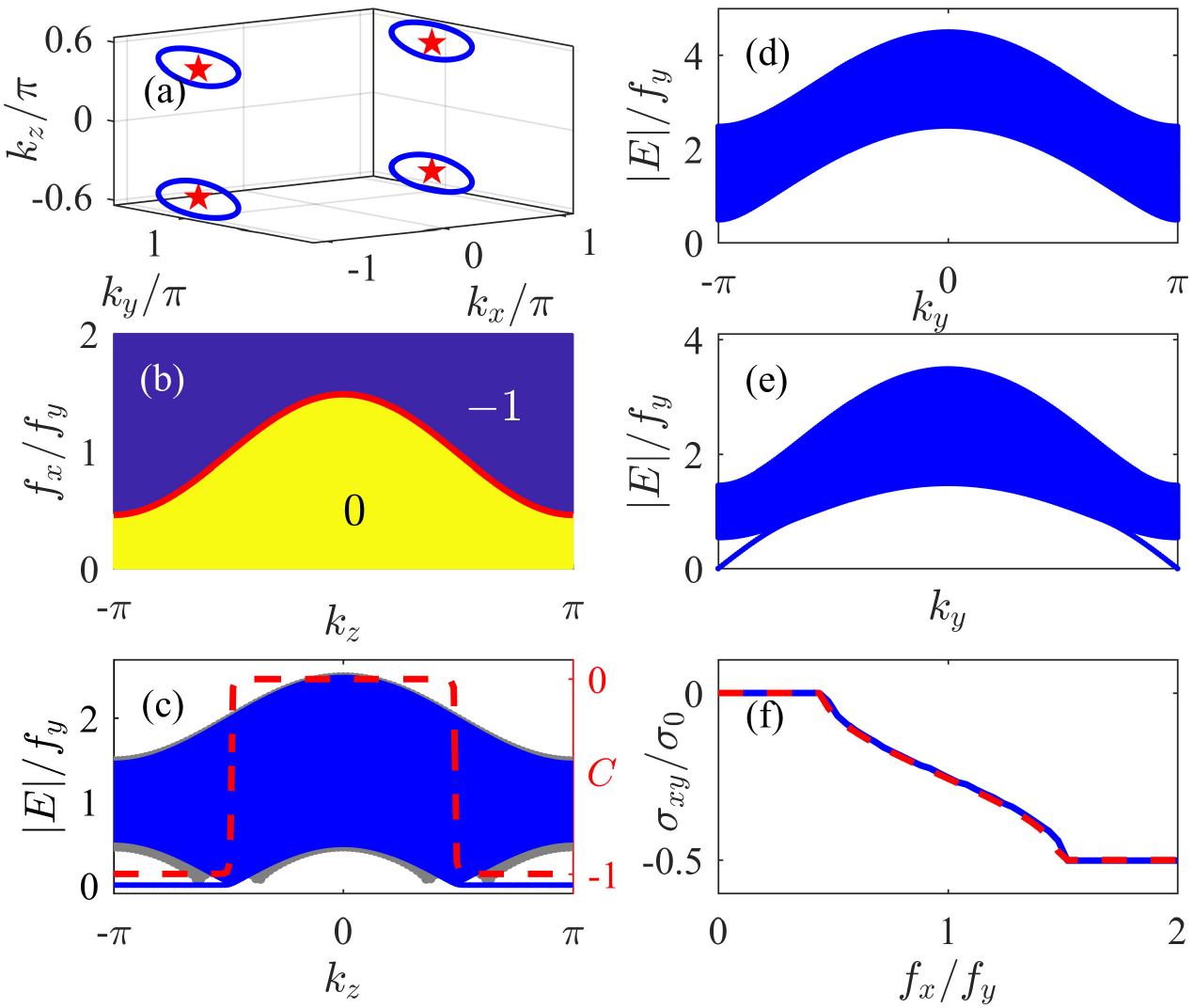}
\caption{(a) Exceptional rings in CBZ (blue solid lines) and points in GBZ (red stars). (b) Phase diagram described by $C(k_z)$. The red solid line is obtained by setting Eq. \eqref{eign} as zero. (c) Energy spectrum under $x$-direction OBC for $k_y=\pi$ and Chern number when $f_x=f_y$. The one under the PBC is shown by the gray area. Energy spectra under the $x$-direction OBC when $k_z=0$ (d) and $\pi$ (e). (f) Hall conductivity from Eqs. \eqref{hcd} (blue solid line) and \eqref{sigmaxy} (red dashed line), where $\sigma_0=e^2/h$. We use $f_z=0.5f_y$, $m=2f_y$, and $\gamma=0.4f_y$.     } \label{Fig1}
\end{figure}

Its Hermitian counterpart is a nodal-point semimetal because the bands touch at discrete points ${\bf k}=[p,p',\arccos(-{m+f_x \cos p+f_y\cos p'\over f_z})]$, with $p$ and $p'$ being $0$ or $\pm\pi$, when $\gamma=0$. Equation \eqref{Hamt} reveals that the non-Hermitian term converts the points into exceptional rings in the $k_x\equiv p=0$ and $\pm\pi$ planes satisfying $(F_{k_y,k_z}+f_xe^{ip})^2+f_y^2\sin^2k_y-\gamma^2/4=0$ [see the solid lines in Fig. \ref{Fig1}(a)]. Thus, being consistent with Refs. \cite{PhysRevB.97.075128,Cerjan2019,PhysRevB.100.245205,PhysRevResearch.2.043268,PhysRevLett.127.196801,PhysRevA.103.L051101}, we really find in the conventional Brillouin zone (CBZ) that the system becomes a Weyl-exceptional-ring semimetal. To verify if this is still true under the OBC, we introduce a generalized BZ (GBZ) formed by $\tilde{\bf k}=(k_x-i\ln r,k_y,k_z)$ with $r=|(F_{k_y,k_z}-\gamma/2)/(F_{k_y,k_z}+\gamma/2)|^{1/2}$ \cite{supplementary}. It recovers the BBC and permits us to correctly describe the energy spectrum under the OBC. Then Eq. \eqref{Hamt} becomes
\begin{equation}\mathcal{H}(\tilde{\bf k})=\left(\begin{array}{cc}
f_y\sin k_y & F_{k_y,k_z}+{\gamma\over 2}+{f_x\beta^{-1}} \\
F_{k_y,k_z}-{\gamma\over 2}+f_x\beta  & -f_y\sin k_y \\
\end{array}\right) \label{genwa}\end{equation} with $\beta\equiv e^{i\tilde{k}_x}=re^{ik_x}$. Its eigen energies are \begin{eqnarray}
E^2&=&F_{k_y,k_z}^2+f^2_x-\gamma^2/4+f^2_y\sin^2{k_y}\nonumber\\
&&+(F_{k_y,k_z}-{\gamma\over2})f_x\beta^{-1}+(F_{k_y,k_z}+{\gamma\over2})f_x\beta.\label{eign}\end{eqnarray} We find that the bands still touch at the discrete values ${\bf k}=[p,p',\pm\arccos{\pm\sqrt{f_x^2+\gamma^2/4}-m-f_y\cos p'\over f_z}] $, as shown by the stars in Fig. \ref{Fig1}(a). Different from the result of Eq. \eqref{Hamt} under the PBC, the system is still a Weyl-exceptional-point semimetal. It demonstrates the leading role of the skin effect in determining the feature of NHWSs.

The GBZ permits us to topologically characterize the NHWS, which is sliced into a family of two-dimensional (2D) topological and normal insulators parameterized by $k_z$. The $k_z$-dependent Chern number is defined in the GBZ as
\begin{equation}
C(k_z)=\frac{1}{4\pi}\int \underline{\mathbf{h}}\cdot\big(\partial_{\tilde{k}_x} \underline{\mathbf{h}}\times \partial_{k_y} \underline{\mathbf{h}}\big) d\tilde{k}_xdk_y,
\end{equation}
where $\underline{\mathbf{h}}={\mathbf{h}}/h$ with $\bf{h}=\text{Tr}[\mathcal{H}(\tilde{k}){\pmb \sigma}]$ and $h=\sqrt{\mathbf{h}\cdot\mathbf{h}}$ \cite{PhysRevB.100.081104}. We plot in Fig. \ref{Fig1}(b) the phase diagram by calculating $C(k_z)$. A topologically non-trivial phase signified by $C(k_z)=-1$ is formed after the EPs obtainable by setting Eq. \eqref{eign} as zero. When $f_x<0.4f_y$, $C(k_z)=0$ for all $k_z$ and the system is a three-dimensional (3D) normal insulator. When $f_x>1.5f_y$, $C(k_z)=-1$ for all $k_z$ and the system is a 3D topological insulator. A NHWS is formed when $f_x\in[0.4,1.5]f_y$. This is confirmed by the energy spectrum under the $x$-direction OBC. We see from Fig. \ref{Fig1}(c) that, exhibiting a severe deviation from the one under the OBC due to the skin effect, the energy spectrum under the OBC is well described by $C(k_z)$. Two Weyl EPs are present where $C(k_z)$ jumps between $0$ and $-1$. A line connecting the two EPs called Fermi arc is formed in the regime of $C(k_z)=-1$. The regime without the Fermi arc is a normal insulator [Fig. \ref{Fig1}(d)], while the one with the Fermi arc is a 2D Chern insulator [Fig. \ref{Fig1}(e)]. Thus, a complete topological description to the NHWS is established.

The GBZ also permits us to calculate the Hall conductivity. Inspired by its 2D form \cite{PhysRevB.100.081104}, we define it as
\begin{eqnarray}
\sigma_{xy}=\frac{-\pi e^2}{2h}\int \frac{d^3\tilde{\bf k}}{(2\pi)^3}\text{Re}\big[\mathbf{\underline{h}}\cdot\big(\partial_{\tilde{k}_x} \underline{\mathbf{h}} \times\partial_{k_y} \underline{\mathbf{h}}\big)\big]\text{sgn}[\text{Re}(h)].~~~~\label{hcd}
\end{eqnarray}
It is verified that $\text{sgn}[\text{Re}(h)]=-1$ for the lower band of our system, which reduces Eq. \eqref{hcd} into $\sigma_{xy}=\frac{e^2}{4\pi h}\int dk_zC(k_z)$. If the system is a 3D normal insulator, then $C(k_z)=0$ for all $k_z$ and thus $\sigma_{xy}=0$. If it is a 3D topological insulator, then $C(k_z)=-1$ for all $k_z$ and thus $\sigma_{xy}=-e^2/(2h)$. If it is a Weyl semimetal, then $\sigma_{xy}\in (-e^2/(2h),0)$ is governed by the length of the Fermi arc equal to
\begin{equation}
\sigma_{xy}=\frac{e^2}{2h}[\frac{1}{\pi}\arccos\frac{\pm\sqrt{f^2_x+\gamma^2/4}-m-f_y\cos p'}{f_z}-1].\label{sigmaxy}
\end{equation}
Figure \ref{Fig1}(f) shows $\sigma_{xy}$ in different $f_x$. It exhibits two obvious quantised plateaus in the three typical regimes, which corresponds to the phases in Fig. \ref{Fig1}(b), i.e., the 3D normal insulator, the NHWS, and the 3D topological insulator. Thus, $\sigma_{xy}$ as an experimental observable can be used to characterize the phase transition in our system.

\section{Floquet engineering to NHWS}
For creating more exotic NHWSs, we make $f_x$ periodically change as
\begin{equation}
f_x(t) =\begin{cases}q_1 f , ~t\in\lbrack zT, zT+T_1),\\q_2f, ~t\in\lbrack zT+T_1, (z+1)T),\end{cases} \label{procotol}
\end{equation}
where $z\in \mathbb{Z}$ and $T=T_1+T_2$ is the period. The Hamiltonian periodically takes two pairwise forms $\hat{H}_1$ and $\hat{H}_2$ within their respective time durations $T_1$ and $T_2$. Such a type of driving has been used in generating a time crystal \cite{Choi2017,Zhang2017}. The periodic system does not have an energy spectrum because the energy is not conserved. According to Floquet theorem, the one-period evolution operator $\hat{U}_T=e^{-i\hat{{H}}_2T_2}e^{-i\hat{{H}}_1T_1}$ defines an effective Hamiltonian $\hat{H}_\text{eff}=\frac{i}{T}\ln \hat{U}_T$ whose eigenvalues are called quasienergies. The topological properties of periodic systems are defined in such quasienergy spectrum. Applying Floquet theorem in $\mathcal{H}_j={\bf h}_j\cdot\pmb{\sigma}$ ($j=1,2$), we obtain $\mathcal{H}_\text{eff}(\mathbf{k})=\mathbf{h}_\text{eff}(\mathbf{k})\cdot{\pmb\sigma}=\frac{i}{T}\ln[e^{-i\mathcal{H}_2(\mathbf{k})T_2}e^{-i\mathcal{H}_1(\mathbf{k})T_1}]$ with $\mathbf{h}_\text{eff}(\mathbf{k})=-\arccos (\epsilon) \underline{\mathbf{r}}/T$ and \cite{PhysRevB.102.041119}
\begin{eqnarray}
\epsilon&=&\cos(T_1 h_1)\cos(T_2 h_2)-\underline{\mathbf{h}}_1\cdot\underline{\mathbf{h}}_2\sin(T_1h_1)\sin(T_2h_2),~~~\label{epsl}\\
\mathbf{r}&=& \underline{\mathbf{h}}_1\times\underline{\mathbf{h}}_2\sin(T_1h_1)\sin(T_2h_2)-\underline{\mathbf{h}}_2\cos(T_1h_1)\nonumber\\ &&\times\sin(T_2h_2)-\underline{\mathbf{h}}_1\cos(T_2h_2)\sin(T_1h_1).\label{varepsilon}
\end{eqnarray}
The bands touch at zero and $\pi/T$ when $\epsilon=1$ and $-1$, respectively. Thus, the EPs fulfill either
\begin{equation}
T_jh_j=n_j\pi,~n_j\in \mathbb{Z}, \label{gen}
\end{equation}\vspace{-1.cm}
\begin{subequations}
  \label{hh1}
    \begin{empheq}
     [left={\text{or}~\empheqlbrace\,}]{align}
      & \underline{\mathbf{h}}_1\cdot\underline{\mathbf{h}}_2=\pm1,  \label{eq:1} \\
      & T_1{E}_1\pm T_2{E}_2=n\pi,~n\in\mathbb{Z}.     \label{eq:2}
    \end{empheq}
\end{subequations}
With the help of the GBZ, we can determine the Weyl points of our periodic system. The advantage over the static case is that we can freely manipulate the number and position of the EPs by tuning the driving parameters. This offers us sufficient room to explore the exotic NHWS absent in static systems.

We can prove that the GBZ is not changed by the periodic driving \cite{supplementary}. It is obtained from $\mathbf{h}_j$ in the GBZ that Eq. \eqref{eq:1} is fulfilled when $k_x=p$ and $k_y=p'$ are zero or $\pi$.  Then we obtain from Eq. \eqref{eq:2} that the EPs satisfy
\begin{equation}
\sum_{j=1,2}(\pm1)^jT_j\Big|\sqrt{F^2_{p',k_z}-{\gamma^2/4}}+q_jfe^{ip}\Big|=n_\pm\pi.\label{drcdt}
\end{equation}
Note that condition \eqref{gen} does not cause a $k_z$-dependent topological phase transition and thus does not form Weyl points in our system, which is similar to the periodic Haldane mode \cite{PhysRevB.93.184306}. Equation \eqref{drcdt} supplies us a guideline to control the number and the position of the Weyl EPs by designing the periodic driving.

\begin{figure}[tbp]
\centering
\includegraphics[width=1\columnwidth]{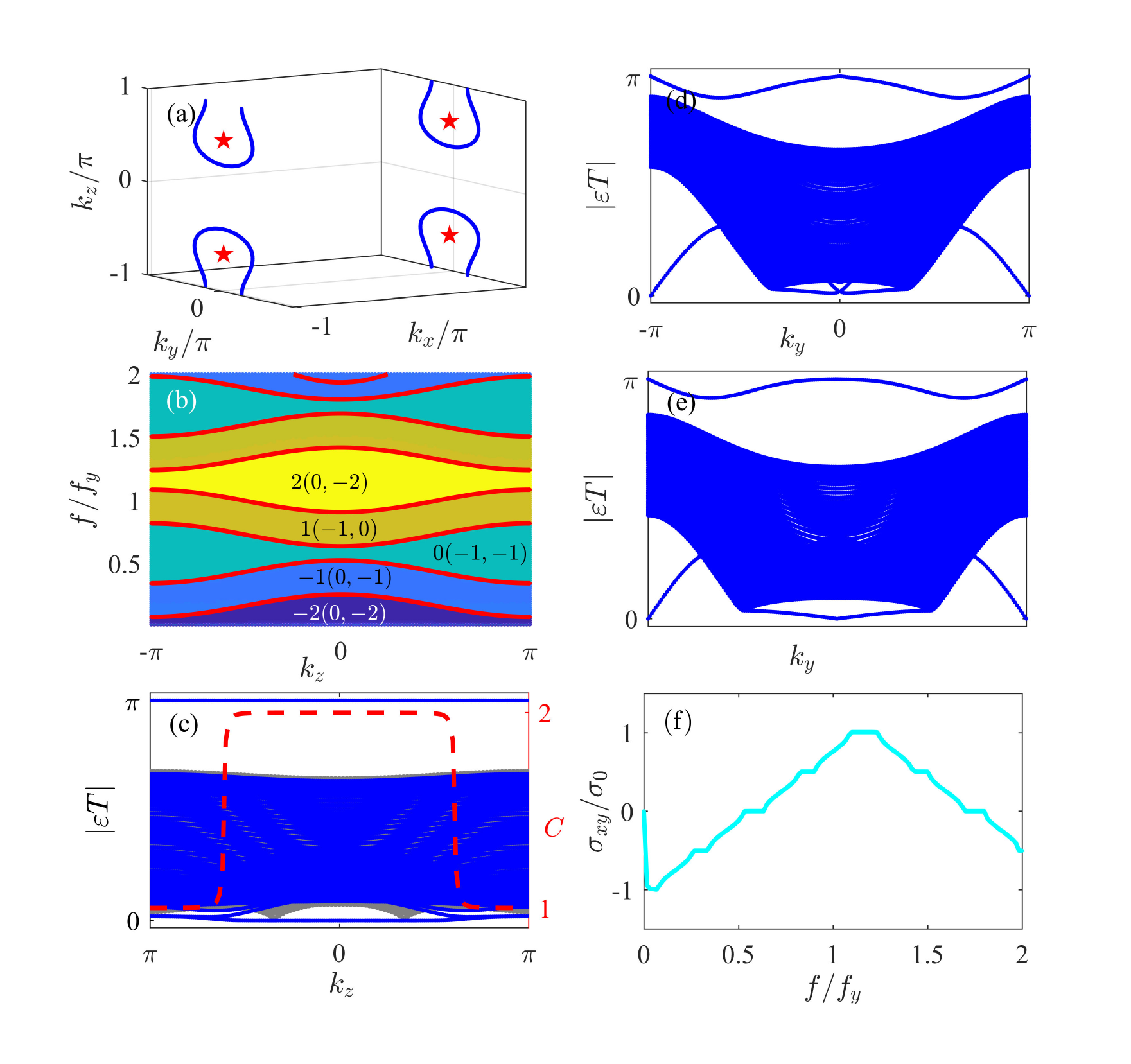}
\caption{(a) Exceptional rings in CBZ (blue solid lines) and points in GBZ (red stars). (b) Phase diagram described by $C(k_z)~[C_0(k_z),C_{\pi/T}(k_z)]$. The red solid lines are from Eq. \eqref{drcdt}. (c) Quasienergy spectrum under $x$-direction OBC for $k_y=0$ and $C(k_z)$ when $f_x=1.3f_y$. The one under the PBC is shown by the gray area. Quasienergy spectra under the $x$-direction OBC when $k_z=\pi$ (d) and $0$ (e). (f) Hall conductivity from Eq. \eqref{hcd}. We use $m=2f_y$, $f_z=0.2f_y$, $\gamma=0.4f_y$, $q_1=1$, $q_2=3.5$, and $T_1=T_2=0.6f^{-1}_y$.  } \label{nl2}
\end{figure}

The periodic driving makes the topological characterization nontrivial. We develop the following scheme. First, a Chern number $C(k_z)$ describing the topology of the overall bands of the zero and $\pi/T$ modes is calculated from $\mathcal{H}_{\rm eff}(\tilde{\mathbf{k}})$. Second, a Chern number describing only the zero mode can be defined by the dynamical way \cite{PhysRevX.3.031005}
\begin{equation}
C_0(k_z)=-C(t=0)-\sum_{j}\mathcal{Q}_j,
\end{equation}
where $C(t=0)$ is the Chern number of the initial Hamiltonian and $\mathcal{Q}_j$ is the topological charge for $j$th EP at time $t\in[0,T]$. The charge $\mathcal{Q}_j$ is defined as $\mathcal{Q}_j=\frac{1}{2\pi}\oint_{\bf \mathcal S}[{\pmb \bigtriangledown}\times {\bf A}(\mathbf{k},t)] \cdot d{\bf S}$, where $ {\bf A}(\mathbf{k},t)=-i\langle u_L(t)\lvert{\pmb \nabla} \lvert u_R(t)\rangle$ is the Berry curvature, ${\mathcal S}$ is a close surface enclosing the EP \cite{PhysRevLett.118.045701}, and $|u_{L,R}(t)\rangle$ are the left and right eigenstates of the evolution operator $U(t)$. Combined with Eq. \eqref{drcdt} for finding the EPs, we can characterize the zero-mode topology. The $\pi/T$-mode topology is characterized by
\begin{equation}
C_{\pi/T}(k_z)=C(k_z)+C_0(k_z).
\end{equation}
The numbers of the zero- and $\pi/T$-mode boundary states are $|C_0(k_z)|$ and $|C_{\pi/T}(k_z)|$ if they have the same chirality \cite{PhysRevB.93.184306}.

To reveal the impact of the skin effect, we plot in Fig. \ref{nl2}(a) the EPs in CBZ and GBZ. The first ones form a ring (see the blue solid line), which depicts a Weyl-exception-ring semimetal, while the second ones still take discrete points (see the red stars), which support a Weyl-exceptional-point semimetal. It reflects the significance of the skin effect in determining the feature of the NHWS in our periodic system. The topologies in both the zero- and $\pi/T$-mode gaps are well described by $C_0(k_z)$ and $C_{\pi/T}(k_z)$ defined in GBZ. The phase diagram in Fig. \ref{nl2}(b) reveal that the more colorful 2D sliced phases with widely tunable $C(k_z)$ than the static case in Fig. \ref{Fig1}(b) are created by the periodic driving. Matching the analytical condition \eqref{drcdt} [see the red solid lines in Fig. \ref{nl2}(b)], the critical points are just the Weyl EPs.
Different from the static case, where the Weyl EPs separate the normal and topological insulators, the ones in our periodic system also separate the topological insulators with different Chern numbers.

An interesting consequence of the Weyl EPs separating the phases with different $C(k_z)$ is that composite topological phases of NHWS and the topological insulator can be formed. Such exotic phases are signified by the coexisting Fermi arcs and the gapless Chiral boundary states. This can be confirmed by the quasienergy spectrum and Chern number. Since two EPs are formed in the $k_y=0$ plane, we plot in Fig. \ref{nl2}(c) the quasienergy spectrum when $k_y=0$ and its Chern number. The Chern number goes from 1 to 2 and back to 1 at the two EPs. A Fermi arc is formed between them in the regime of $C(k_z)=2$. However, there is a pair of gapless chiral boundary states supporting $C(k_z)=1$ in the full $k_z$ regimes [see Figs. \ref{nl2}(d) and \ref{nl2}(e)], which reveals the existence of a topological insulator. It depicts a composite topological phase with coexisting surface Fermi arc for the NHWS and the chiral boundary states for the 3D topological insulator. A similar phase was reported in Ref. \cite{PhysRevB.100.161401}, but it is in Dirac type and Hermitian system. A result of this widely tunable $C(k_z)$ is that the Hall conductivity is much enhanced to exhibit multiple quantized plateaus than the static case [see Fig. \ref{nl2}(f)]. When the system is the 3D topological insulator, $\sigma_{xy}$ exhibits a plateau equaling $-e^2/(2h)$ multiplied by the associated $C(k_z)$. When the system is a NHWS, $\sigma_{xy}$ is proportional to the length of the Fermi arc and its Chern number. Such widely tunable conductivity supplies us a useful way to detect the colorful NHWSs induced by the periodic driving and might inspire a promising application in developing quantum-transport devices.

\begin{figure}[tbp]
\centering
\includegraphics[width=1\columnwidth]{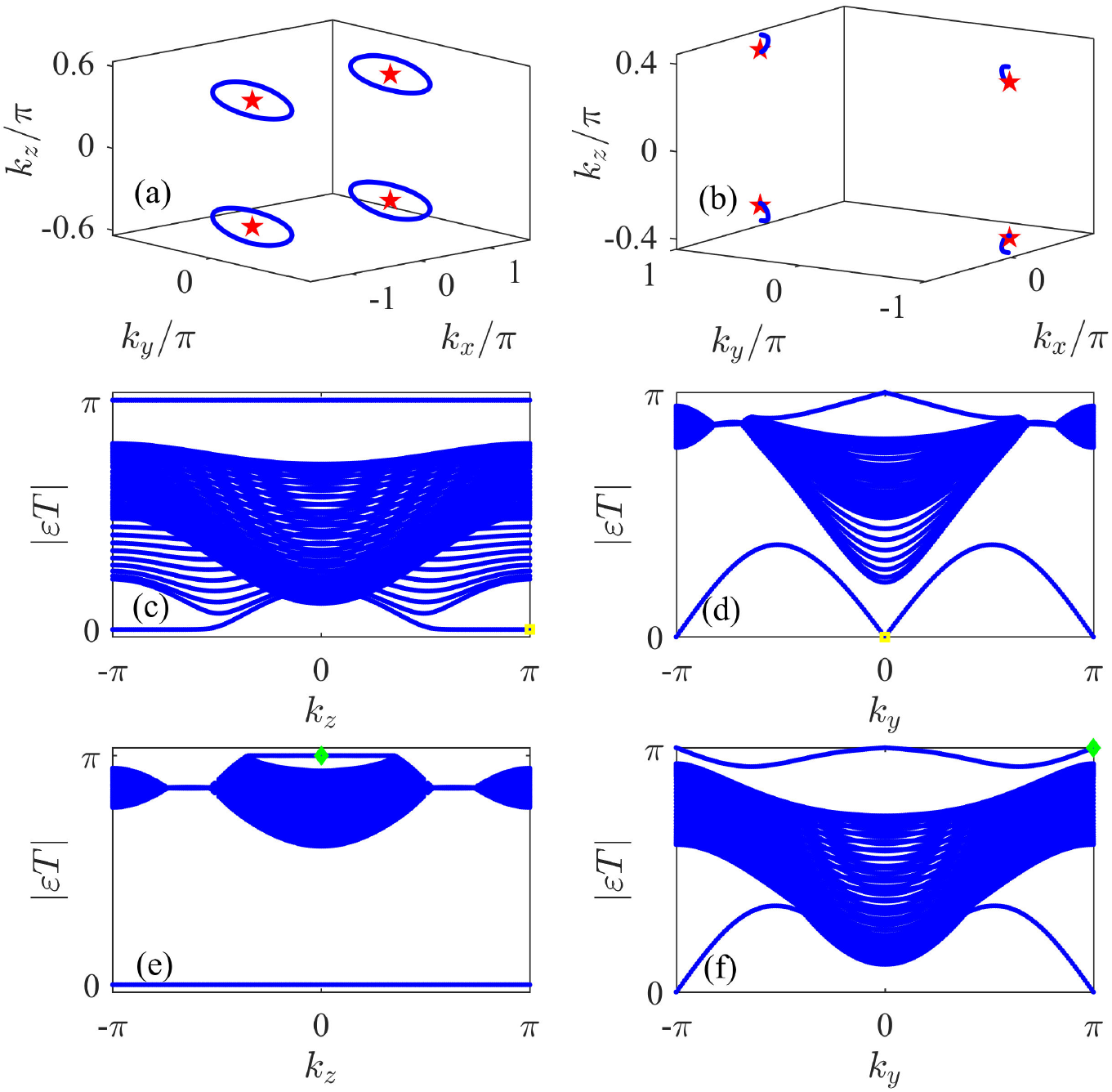}
\caption{Exceptional rings in CBZ (blue solid lines) and points in GBZ (red stars) for the zero-mode (a) and $\pi/T$-mode (b) gaps.  Quasienergy spectra under $x$-direction OBC for $k_y=0$ (c) and $\pi$ in (e), and for $k_z=\pi$ (d) and $0$ (f). We use $m=1.2f_y$, $f_x=f_y$, $f_z=0.5f_y$, $\gamma=0.4f_y$, $q_1=1$, $q_2=3.5$, and $T_1=T_2=0.6f^{-1}_y$. } \label{nl3}
\end{figure}

Another exotic phase induced by the periodic driving is a NHWS with anomalous Fermi arcs. The Fermi arc is generally formed by the crossings of the gapless chiral boundary states of the 2D sliced topological insulators \cite{RevModPhys.90.015001}. We find a counterexample to this. Figures \ref{nl3}(a) and \ref{nl3}(b) indicate that the zero- and $\pi/T$-mode EPs are present at $k_y=0$ and $\pi$ planes, respectively. The quasienergy spectrum for $k_y=0$ in Fig. \ref{nl3}(c) confirms the zero-mode EPs connected by a Fermi arc. To reveal the feature of the Fermi arc, we plot in Fig. \ref{nl3}(d) the quasienergy spectrum when $k_z=\pi$. It is interesting to see that the EP at $k_y=0$ is not an intersecting point of the gapless chiral boundary states, but the one of two gapped bound states. This reveals that the zero-mode Fermi arc occurring at $k_y=0$ in Fig. \ref{nl3}(c) is formed by the intersecting points of these gapped bound states instead of the generally believed gapless chiral boundary states. Such a type of exotic Weyl semimetal is also present in the $\pi/T$ mode. The quasienergy spectrum when $k_y=\pi$ in Fig. \ref{nl3}(e) confirms the EPs in Fig. \ref{nl3}(b) connected by a Fermi arc. The quasienergy spectrum when $k_z=0$ in Fig. \ref{nl3}(f) reveals that this $\pi/T$-mode Fermi arc is also formed by the intersecting point of two gapped bound states. Having not been found before, such exotic NHWS enriches the family of semimetals.

\section{Conclusion}
In summary, we investigated the NHWS and its Floquet engineering in a 3D system. The dominant role of the skin effect in determining the feature of NHWS is revealed in both the static and the periodically driven cases. A complete topological characterization to the NHWS for the two cases is established by introducing the GBZ. Exotic topological phases, e.g., a composite phase of NHWS and topological insulator with the coexisting surface Fermi arc and chiral boundary states, a widely tunable Hall conductivity with multiple quantized plateaus, and a NHWS with anomalous Fermi arc formed by the crossing of the gapped bound state, are created by the periodic driving. Our results indicate that, by supplying a feasible way to engineer novel topological semimetals, the periodic driving is useful in artificially synthesizing the exotic topological phases absent in natural materials. These exotic phases with good controllability might inspire the application and exploration of non-Hermitian topological semimetals.

\section{Acknowledgments}
The work is supported by the National Natural Science Foundation (Grants No. 11875150, No. 11834005, and No. 12047501).
\bibliography{ref}

\begin{thebibliography}{93}%
\makeatletter
\providecommand \@ifxundefined [1]{%
 \@ifx{#1\undefined}
}%
\providecommand \@ifnum [1]{%
 \ifnum #1\expandafter \@firstoftwo
 \else \expandafter \@secondoftwo
 \fi
}%
\providecommand \@ifx [1]{%
 \ifx #1\expandafter \@firstoftwo
 \else \expandafter \@secondoftwo
 \fi
}%
\providecommand \natexlab [1]{#1}%
\providecommand \enquote  [1]{``#1''}%
\providecommand \bibnamefont  [1]{#1}%
\providecommand \bibfnamefont [1]{#1}%
\providecommand \citenamefont [1]{#1}%
\providecommand \href@noop [0]{\@secondoftwo}%
\providecommand \href [0]{\begingroup \@sanitize@url \@href}%
\providecommand \@href[1]{\@@startlink{#1}\@@href}%
\providecommand \@@href[1]{\endgroup#1\@@endlink}%
\providecommand \@sanitize@url [0]{\catcode `\\12\catcode `\$12\catcode
  `\&12\catcode `\#12\catcode `\^12\catcode `\_12\catcode `\%12\relax}%
\providecommand \@@startlink[1]{}%
\providecommand \@@endlink[0]{}%
\providecommand \url  [0]{\begingroup\@sanitize@url \@url }%
\providecommand \@url [1]{\endgroup\@href {#1}{\urlprefix }}%
\providecommand \urlprefix  [0]{URL }%
\providecommand \Eprint [0]{\href }%
\providecommand \doibase [0]{http://dx.doi.org/}%
\providecommand \selectlanguage [0]{\@gobble}%
\providecommand \bibinfo  [0]{\@secondoftwo}%
\providecommand \bibfield  [0]{\@secondoftwo}%
\providecommand \translation [1]{[#1]}%
\providecommand \BibitemOpen [0]{}%
\providecommand \bibitemStop [0]{}%
\providecommand \bibitemNoStop [0]{.\EOS\space}%
\providecommand \EOS [0]{\spacefactor3000\relax}%
\providecommand \BibitemShut  [1]{\csname bibitem#1\endcsname}%
\let\auto@bib@innerbib\@empty
%</preamble>
\bibitem [{\citenamefont {Lee}(2016)}]{PhysRevLett.116.133903}%
  \BibitemOpen
  \bibfield  {author} {\bibinfo {author} {\bibfnamefont {T.~E.}\ \bibnamefont
  {Lee}},\ }\href {\doibase 10.1103/PhysRevLett.116.133903} {\bibfield
  {journal} {\bibinfo  {journal} {Phys. Rev. Lett.}\ }\textbf {\bibinfo
  {volume} {116}},\ \bibinfo {pages} {133903} (\bibinfo {year}
  {2016})}\BibitemShut {NoStop}%
\bibitem [{\citenamefont {Yin}\ \emph {et~al.}(2018)\citenamefont {Yin},
  \citenamefont {Jiang}, \citenamefont {Li}, \citenamefont {L\"u},\ and\
  \citenamefont {Chen}}]{PhysRevA.97.052115}%
  \BibitemOpen
  \bibfield  {author} {\bibinfo {author} {\bibfnamefont {C.}~\bibnamefont
  {Yin}}, \bibinfo {author} {\bibfnamefont {H.}~\bibnamefont {Jiang}}, \bibinfo
  {author} {\bibfnamefont {L.}~\bibnamefont {Li}}, \bibinfo {author}
  {\bibfnamefont {R.}~\bibnamefont {L\"u}}, \ and\ \bibinfo {author}
  {\bibfnamefont {S.}~\bibnamefont {Chen}},\ }\href {\doibase
  10.1103/PhysRevA.97.052115} {\bibfield  {journal} {\bibinfo  {journal} {Phys.
  Rev. A}\ }\textbf {\bibinfo {volume} {97}},\ \bibinfo {pages} {052115}
  (\bibinfo {year} {2018})}\BibitemShut {NoStop}%
\bibitem [{\citenamefont {Yao}\ and\ \citenamefont
  {Wang}(2018)}]{PhysRevLett.121.086803}%
  \BibitemOpen
  \bibfield  {author} {\bibinfo {author} {\bibfnamefont {S.}~\bibnamefont
  {Yao}}\ and\ \bibinfo {author} {\bibfnamefont {Z.}~\bibnamefont {Wang}},\
  }\href {\doibase 10.1103/PhysRevLett.121.086803} {\bibfield  {journal}
  {\bibinfo  {journal} {Phys. Rev. Lett.}\ }\textbf {\bibinfo {volume} {121}},\
  \bibinfo {pages} {086803} (\bibinfo {year} {2018})}\BibitemShut {NoStop}%
\bibitem [{\citenamefont {Kunst}\ \emph {et~al.}(2018)\citenamefont {Kunst},
  \citenamefont {Edvardsson}, \citenamefont {Budich},\ and\ \citenamefont
  {Bergholtz}}]{PhysRevLett.121.026808}%
  \BibitemOpen
  \bibfield  {author} {\bibinfo {author} {\bibfnamefont {F.~K.}\ \bibnamefont
  {Kunst}}, \bibinfo {author} {\bibfnamefont {E.}~\bibnamefont {Edvardsson}},
  \bibinfo {author} {\bibfnamefont {J.~C.}\ \bibnamefont {Budich}}, \ and\
  \bibinfo {author} {\bibfnamefont {E.~J.}\ \bibnamefont {Bergholtz}},\ }\href
  {\doibase 10.1103/PhysRevLett.121.026808} {\bibfield  {journal} {\bibinfo
  {journal} {Phys. Rev. Lett.}\ }\textbf {\bibinfo {volume} {121}},\ \bibinfo
  {pages} {026808} (\bibinfo {year} {2018})}\BibitemShut {NoStop}%
\bibitem [{\citenamefont {Yokomizo}\ and\ \citenamefont
  {Murakami}(2019)}]{PhysRevLett.123.066404}%
  \BibitemOpen
  \bibfield  {author} {\bibinfo {author} {\bibfnamefont {K.}~\bibnamefont
  {Yokomizo}}\ and\ \bibinfo {author} {\bibfnamefont {S.}~\bibnamefont
  {Murakami}},\ }\href {\doibase 10.1103/PhysRevLett.123.066404} {\bibfield
  {journal} {\bibinfo  {journal} {Phys. Rev. Lett.}\ }\textbf {\bibinfo
  {volume} {123}},\ \bibinfo {pages} {066404} (\bibinfo {year}
  {2019})}\BibitemShut {NoStop}%
\bibitem [{\citenamefont {Jin}\ and\ \citenamefont
  {Song}(2019)}]{PhysRevB.99.081103}%
  \BibitemOpen
  \bibfield  {author} {\bibinfo {author} {\bibfnamefont {L.}~\bibnamefont
  {Jin}}\ and\ \bibinfo {author} {\bibfnamefont {Z.}~\bibnamefont {Song}},\
  }\href {\doibase 10.1103/PhysRevB.99.081103} {\bibfield  {journal} {\bibinfo
  {journal} {Phys. Rev. B}\ }\textbf {\bibinfo {volume} {99}},\ \bibinfo
  {pages} {081103(R)} (\bibinfo {year} {2019})}\BibitemShut {NoStop}%
\bibitem [{\citenamefont {Kawabata}\ \emph
  {et~al.}(2019{\natexlab{a}})\citenamefont {Kawabata}, \citenamefont
  {Shiozaki}, \citenamefont {Ueda},\ and\ \citenamefont
  {Sato}}]{PhysRevX.9.041015}%
  \BibitemOpen
  \bibfield  {author} {\bibinfo {author} {\bibfnamefont {K.}~\bibnamefont
  {Kawabata}}, \bibinfo {author} {\bibfnamefont {K.}~\bibnamefont {Shiozaki}},
  \bibinfo {author} {\bibfnamefont {M.}~\bibnamefont {Ueda}}, \ and\ \bibinfo
  {author} {\bibfnamefont {M.}~\bibnamefont {Sato}},\ }\href {\doibase
  10.1103/PhysRevX.9.041015} {\bibfield  {journal} {\bibinfo  {journal} {Phys.
  Rev. X}\ }\textbf {\bibinfo {volume} {9}},\ \bibinfo {pages} {041015}
  (\bibinfo {year} {2019}{\natexlab{a}})}\BibitemShut {NoStop}%
\bibitem [{\citenamefont {Borgnia}\ \emph {et~al.}(2020)\citenamefont
  {Borgnia}, \citenamefont {Kruchkov},\ and\ \citenamefont
  {Slager}}]{PhysRevLett.124.056802}%
  \BibitemOpen
  \bibfield  {author} {\bibinfo {author} {\bibfnamefont {D.~S.}\ \bibnamefont
  {Borgnia}}, \bibinfo {author} {\bibfnamefont {A.~J.}\ \bibnamefont
  {Kruchkov}}, \ and\ \bibinfo {author} {\bibfnamefont {R.-J.}\ \bibnamefont
  {Slager}},\ }\href {\doibase 10.1103/PhysRevLett.124.056802} {\bibfield
  {journal} {\bibinfo  {journal} {Phys. Rev. Lett.}\ }\textbf {\bibinfo
  {volume} {124}},\ \bibinfo {pages} {056802} (\bibinfo {year}
  {2020})}\BibitemShut {NoStop}%
\bibitem [{\citenamefont {Zhang}\ \emph
  {et~al.}(2020{\natexlab{a}})\citenamefont {Zhang}, \citenamefont {Yang},\
  and\ \citenamefont {Fang}}]{PhysRevLett.125.126402}%
  \BibitemOpen
  \bibfield  {author} {\bibinfo {author} {\bibfnamefont {K.}~\bibnamefont
  {Zhang}}, \bibinfo {author} {\bibfnamefont {Z.}~\bibnamefont {Yang}}, \ and\
  \bibinfo {author} {\bibfnamefont {C.}~\bibnamefont {Fang}},\ }\href {\doibase
  10.1103/PhysRevLett.125.126402} {\bibfield  {journal} {\bibinfo  {journal}
  {Phys. Rev. Lett.}\ }\textbf {\bibinfo {volume} {125}},\ \bibinfo {pages}
  {126402} (\bibinfo {year} {2020}{\natexlab{a}})}\BibitemShut {NoStop}%
\bibitem [{\citenamefont {Yi}\ and\ \citenamefont
  {Yang}(2020)}]{PhysRevLett.125.186802}%
  \BibitemOpen
  \bibfield  {author} {\bibinfo {author} {\bibfnamefont {Y.}~\bibnamefont
  {Yi}}\ and\ \bibinfo {author} {\bibfnamefont {Z.}~\bibnamefont {Yang}},\
  }\href {\doibase 10.1103/PhysRevLett.125.186802} {\bibfield  {journal}
  {\bibinfo  {journal} {Phys. Rev. Lett.}\ }\textbf {\bibinfo {volume} {125}},\
  \bibinfo {pages} {186802} (\bibinfo {year} {2020})}\BibitemShut {NoStop}%
\bibitem [{\citenamefont {Leykam}\ \emph {et~al.}(2017)\citenamefont {Leykam},
  \citenamefont {Bliokh}, \citenamefont {Huang}, \citenamefont {Chong},\ and\
  \citenamefont {Nori}}]{PhysRevLett.118.040401}%
  \BibitemOpen
  \bibfield  {author} {\bibinfo {author} {\bibfnamefont {D.}~\bibnamefont
  {Leykam}}, \bibinfo {author} {\bibfnamefont {K.~Y.}\ \bibnamefont {Bliokh}},
  \bibinfo {author} {\bibfnamefont {C.}~\bibnamefont {Huang}}, \bibinfo
  {author} {\bibfnamefont {Y.~D.}\ \bibnamefont {Chong}}, \ and\ \bibinfo
  {author} {\bibfnamefont {F.}~\bibnamefont {Nori}},\ }\href {\doibase
  10.1103/PhysRevLett.118.040401} {\bibfield  {journal} {\bibinfo  {journal}
  {Phys. Rev. Lett.}\ }\textbf {\bibinfo {volume} {118}},\ \bibinfo {pages}
  {040401} (\bibinfo {year} {2017})}\BibitemShut {NoStop}%
\bibitem [{\citenamefont {Shen}\ \emph {et~al.}(2018)\citenamefont {Shen},
  \citenamefont {Zhen},\ and\ \citenamefont {Fu}}]{PhysRevLett.120.146402}%
  \BibitemOpen
  \bibfield  {author} {\bibinfo {author} {\bibfnamefont {H.}~\bibnamefont
  {Shen}}, \bibinfo {author} {\bibfnamefont {B.}~\bibnamefont {Zhen}}, \ and\
  \bibinfo {author} {\bibfnamefont {L.}~\bibnamefont {Fu}},\ }\href {\doibase
  10.1103/PhysRevLett.120.146402} {\bibfield  {journal} {\bibinfo  {journal}
  {Phys. Rev. Lett.}\ }\textbf {\bibinfo {volume} {120}},\ \bibinfo {pages}
  {146402} (\bibinfo {year} {2018})}\BibitemShut {NoStop}%
\bibitem [{\citenamefont {Song}\ \emph {et~al.}(2020)\citenamefont {Song},
  \citenamefont {Sun}, \citenamefont {Dutt}, \citenamefont {Minkov},
  \citenamefont {Wojcik}, \citenamefont {Wang}, \citenamefont {Williamson},
  \citenamefont {Orenstein},\ and\ \citenamefont
  {Fan}}]{PhysRevLett.125.033603}%
  \BibitemOpen
  \bibfield  {author} {\bibinfo {author} {\bibfnamefont {A.~Y.}\ \bibnamefont
  {Song}}, \bibinfo {author} {\bibfnamefont {X.-Q.}\ \bibnamefont {Sun}},
  \bibinfo {author} {\bibfnamefont {A.}~\bibnamefont {Dutt}}, \bibinfo {author}
  {\bibfnamefont {M.}~\bibnamefont {Minkov}}, \bibinfo {author} {\bibfnamefont
  {C.}~\bibnamefont {Wojcik}}, \bibinfo {author} {\bibfnamefont
  {H.}~\bibnamefont {Wang}}, \bibinfo {author} {\bibfnamefont {I.~A.~D.}\
  \bibnamefont {Williamson}}, \bibinfo {author} {\bibfnamefont
  {M.}~\bibnamefont {Orenstein}}, \ and\ \bibinfo {author} {\bibfnamefont
  {S.}~\bibnamefont {Fan}},\ }\href {\doibase 10.1103/PhysRevLett.125.033603}
  {\bibfield  {journal} {\bibinfo  {journal} {Phys. Rev. Lett.}\ }\textbf
  {\bibinfo {volume} {125}},\ \bibinfo {pages} {033603} (\bibinfo {year}
  {2020})}\BibitemShut {NoStop}%
\bibitem [{\citenamefont {Yang}\ \emph
  {et~al.}(2020{\natexlab{a}})\citenamefont {Yang}, \citenamefont {Zhang},
  \citenamefont {Fang},\ and\ \citenamefont {Hu}}]{PhysRevLett.125.226402}%
  \BibitemOpen
  \bibfield  {author} {\bibinfo {author} {\bibfnamefont {Z.}~\bibnamefont
  {Yang}}, \bibinfo {author} {\bibfnamefont {K.}~\bibnamefont {Zhang}},
  \bibinfo {author} {\bibfnamefont {C.}~\bibnamefont {Fang}}, \ and\ \bibinfo
  {author} {\bibfnamefont {J.}~\bibnamefont {Hu}},\ }\href {\doibase
  10.1103/PhysRevLett.125.226402} {\bibfield  {journal} {\bibinfo  {journal}
  {Phys. Rev. Lett.}\ }\textbf {\bibinfo {volume} {125}},\ \bibinfo {pages}
  {226402} (\bibinfo {year} {2020}{\natexlab{a}})}\BibitemShut {NoStop}%
\bibitem [{\citenamefont {Longhi}(2019)}]{PhysRevLett.122.237601}%
  \BibitemOpen
  \bibfield  {author} {\bibinfo {author} {\bibfnamefont {S.}~\bibnamefont
  {Longhi}},\ }\href {\doibase 10.1103/PhysRevLett.122.237601} {\bibfield
  {journal} {\bibinfo  {journal} {Phys. Rev. Lett.}\ }\textbf {\bibinfo
  {volume} {122}},\ \bibinfo {pages} {237601} (\bibinfo {year}
  {2019})}\BibitemShut {NoStop}%
\bibitem [{\citenamefont {Gao}\ \emph {et~al.}(2020)\citenamefont {Gao},
  \citenamefont {Willatzen},\ and\ \citenamefont
  {Christensen}}]{PhysRevLett.125.206402}%
  \BibitemOpen
  \bibfield  {author} {\bibinfo {author} {\bibfnamefont {P.}~\bibnamefont
  {Gao}}, \bibinfo {author} {\bibfnamefont {M.}~\bibnamefont {Willatzen}}, \
  and\ \bibinfo {author} {\bibfnamefont {J.}~\bibnamefont {Christensen}},\
  }\href {\doibase 10.1103/PhysRevLett.125.206402} {\bibfield  {journal}
  {\bibinfo  {journal} {Phys. Rev. Lett.}\ }\textbf {\bibinfo {volume} {125}},\
  \bibinfo {pages} {206402} (\bibinfo {year} {2020})}\BibitemShut {NoStop}%
\bibitem [{\citenamefont {Zeuner}\ \emph {et~al.}(2015)\citenamefont {Zeuner},
  \citenamefont {Rechtsman}, \citenamefont {Plotnik}, \citenamefont {Lumer},
  \citenamefont {Nolte}, \citenamefont {Rudner}, \citenamefont {Segev},\ and\
  \citenamefont {Szameit}}]{PhysRevLett.115.040402}%
  \BibitemOpen
  \bibfield  {author} {\bibinfo {author} {\bibfnamefont {J.~M.}\ \bibnamefont
  {Zeuner}}, \bibinfo {author} {\bibfnamefont {M.~C.}\ \bibnamefont
  {Rechtsman}}, \bibinfo {author} {\bibfnamefont {Y.}~\bibnamefont {Plotnik}},
  \bibinfo {author} {\bibfnamefont {Y.}~\bibnamefont {Lumer}}, \bibinfo
  {author} {\bibfnamefont {S.}~\bibnamefont {Nolte}}, \bibinfo {author}
  {\bibfnamefont {M.~S.}\ \bibnamefont {Rudner}}, \bibinfo {author}
  {\bibfnamefont {M.}~\bibnamefont {Segev}}, \ and\ \bibinfo {author}
  {\bibfnamefont {A.}~\bibnamefont {Szameit}},\ }\href {\doibase
  10.1103/PhysRevLett.115.040402} {\bibfield  {journal} {\bibinfo  {journal}
  {Phys. Rev. Lett.}\ }\textbf {\bibinfo {volume} {115}},\ \bibinfo {pages}
  {040402} (\bibinfo {year} {2015})}\BibitemShut {NoStop}%
\bibitem [{\citenamefont {Ding}\ \emph {et~al.}(2016)\citenamefont {Ding},
  \citenamefont {Ma}, \citenamefont {Xiao}, \citenamefont {Zhang},\ and\
  \citenamefont {Chan}}]{PhysRevX.6.021007}%
  \BibitemOpen
  \bibfield  {author} {\bibinfo {author} {\bibfnamefont {K.}~\bibnamefont
  {Ding}}, \bibinfo {author} {\bibfnamefont {G.}~\bibnamefont {Ma}}, \bibinfo
  {author} {\bibfnamefont {M.}~\bibnamefont {Xiao}}, \bibinfo {author}
  {\bibfnamefont {Z.~Q.}\ \bibnamefont {Zhang}}, \ and\ \bibinfo {author}
  {\bibfnamefont {C.~T.}\ \bibnamefont {Chan}},\ }\href {\doibase
  10.1103/PhysRevX.6.021007} {\bibfield  {journal} {\bibinfo  {journal} {Phys.
  Rev. X}\ }\textbf {\bibinfo {volume} {6}},\ \bibinfo {pages} {021007}
  (\bibinfo {year} {2016})}\BibitemShut {NoStop}%
\bibitem [{\citenamefont {Zhu}\ \emph {et~al.}(2018)\citenamefont {Zhu},
  \citenamefont {Fang}, \citenamefont {Li}, \citenamefont {Sun}, \citenamefont
  {Li}, \citenamefont {Jing},\ and\ \citenamefont
  {Chen}}]{PhysRevLett.121.124501}%
  \BibitemOpen
  \bibfield  {author} {\bibinfo {author} {\bibfnamefont {W.}~\bibnamefont
  {Zhu}}, \bibinfo {author} {\bibfnamefont {X.}~\bibnamefont {Fang}}, \bibinfo
  {author} {\bibfnamefont {D.}~\bibnamefont {Li}}, \bibinfo {author}
  {\bibfnamefont {Y.}~\bibnamefont {Sun}}, \bibinfo {author} {\bibfnamefont
  {Y.}~\bibnamefont {Li}}, \bibinfo {author} {\bibfnamefont {Y.}~\bibnamefont
  {Jing}}, \ and\ \bibinfo {author} {\bibfnamefont {H.}~\bibnamefont {Chen}},\
  }\href {\doibase 10.1103/PhysRevLett.121.124501} {\bibfield  {journal}
  {\bibinfo  {journal} {Phys. Rev. Lett.}\ }\textbf {\bibinfo {volume} {121}},\
  \bibinfo {pages} {124501} (\bibinfo {year} {2018})}\BibitemShut {NoStop}%
\bibitem [{\citenamefont {Xiao}\ \emph {et~al.}(2020)\citenamefont {Xiao},
  \citenamefont {Deng}, \citenamefont {Wang}, \citenamefont {Zhu},
  \citenamefont {Wang}, \citenamefont {Yi},\ and\ \citenamefont
  {Xue}}]{Xiao2020}%
  \BibitemOpen
  \bibfield  {author} {\bibinfo {author} {\bibfnamefont {L.}~\bibnamefont
  {Xiao}}, \bibinfo {author} {\bibfnamefont {T.}~\bibnamefont {Deng}}, \bibinfo
  {author} {\bibfnamefont {K.}~\bibnamefont {Wang}}, \bibinfo {author}
  {\bibfnamefont {G.}~\bibnamefont {Zhu}}, \bibinfo {author} {\bibfnamefont
  {Z.}~\bibnamefont {Wang}}, \bibinfo {author} {\bibfnamefont {W.}~\bibnamefont
  {Yi}}, \ and\ \bibinfo {author} {\bibfnamefont {P.}~\bibnamefont {Xue}},\
  }\href {\doibase 10.1038/s41567-020-0836-6} {\bibfield  {journal} {\bibinfo
  {journal} {Nature Physics}\ }\textbf {\bibinfo {volume} {16}},\ \bibinfo
  {pages} {761} (\bibinfo {year} {2020})}\BibitemShut {NoStop}%
\bibitem [{\citenamefont {Wen}\ \emph {et~al.}(2019)\citenamefont {Wen},
  \citenamefont {Zheng}, \citenamefont {Kong}, \citenamefont {Wei},
  \citenamefont {Xin},\ and\ \citenamefont {Long}}]{PhysRevA.99.062122}%
  \BibitemOpen
  \bibfield  {author} {\bibinfo {author} {\bibfnamefont {J.}~\bibnamefont
  {Wen}}, \bibinfo {author} {\bibfnamefont {C.}~\bibnamefont {Zheng}}, \bibinfo
  {author} {\bibfnamefont {X.}~\bibnamefont {Kong}}, \bibinfo {author}
  {\bibfnamefont {S.}~\bibnamefont {Wei}}, \bibinfo {author} {\bibfnamefont
  {T.}~\bibnamefont {Xin}}, \ and\ \bibinfo {author} {\bibfnamefont
  {G.}~\bibnamefont {Long}},\ }\href {\doibase 10.1103/PhysRevA.99.062122}
  {\bibfield  {journal} {\bibinfo  {journal} {Phys. Rev. A}\ }\textbf {\bibinfo
  {volume} {99}},\ \bibinfo {pages} {062122} (\bibinfo {year}
  {2019})}\BibitemShut {NoStop}%
\bibitem [{\citenamefont {Helbig}\ \emph {et~al.}(2020)\citenamefont {Helbig},
  \citenamefont {Hofmann}, \citenamefont {Imhof}, \citenamefont {Abdelghany},
  \citenamefont {Kiessling}, \citenamefont {Molenkamp}, \citenamefont {Lee},
  \citenamefont {Szameit}, \citenamefont {Greiter},\ and\ \citenamefont
  {Thomale}}]{Helbig2020}%
  \BibitemOpen
  \bibfield  {author} {\bibinfo {author} {\bibfnamefont {T.}~\bibnamefont
  {Helbig}}, \bibinfo {author} {\bibfnamefont {T.}~\bibnamefont {Hofmann}},
  \bibinfo {author} {\bibfnamefont {S.}~\bibnamefont {Imhof}}, \bibinfo
  {author} {\bibfnamefont {M.}~\bibnamefont {Abdelghany}}, \bibinfo {author}
  {\bibfnamefont {T.}~\bibnamefont {Kiessling}}, \bibinfo {author}
  {\bibfnamefont {L.~W.}\ \bibnamefont {Molenkamp}}, \bibinfo {author}
  {\bibfnamefont {C.~H.}\ \bibnamefont {Lee}}, \bibinfo {author} {\bibfnamefont
  {A.}~\bibnamefont {Szameit}}, \bibinfo {author} {\bibfnamefont
  {M.}~\bibnamefont {Greiter}}, \ and\ \bibinfo {author} {\bibfnamefont
  {R.}~\bibnamefont {Thomale}},\ }\href {\doibase 10.1038/s41567-020-0922-9}
  {\bibfield  {journal} {\bibinfo  {journal} {Nature Physics}\ }\textbf
  {\bibinfo {volume} {16}},\ \bibinfo {pages} {747} (\bibinfo {year}
  {2020})}\BibitemShut {NoStop}%
\bibitem [{\citenamefont {Weidemann}\ \emph {et~al.}(2020)\citenamefont
  {Weidemann}, \citenamefont {Kremer}, \citenamefont {Helbig}, \citenamefont
  {Hofmann}, \citenamefont {Stegmaier}, \citenamefont {Greiter}, \citenamefont
  {Thomale},\ and\ \citenamefont {Szameit}}]{Weidemann2020}%
  \BibitemOpen
  \bibfield  {author} {\bibinfo {author} {\bibfnamefont {S.}~\bibnamefont
  {Weidemann}}, \bibinfo {author} {\bibfnamefont {M.}~\bibnamefont {Kremer}},
  \bibinfo {author} {\bibfnamefont {T.}~\bibnamefont {Helbig}}, \bibinfo
  {author} {\bibfnamefont {T.}~\bibnamefont {Hofmann}}, \bibinfo {author}
  {\bibfnamefont {A.}~\bibnamefont {Stegmaier}}, \bibinfo {author}
  {\bibfnamefont {M.}~\bibnamefont {Greiter}}, \bibinfo {author} {\bibfnamefont
  {R.}~\bibnamefont {Thomale}}, \ and\ \bibinfo {author} {\bibfnamefont
  {A.}~\bibnamefont {Szameit}},\ }\href {\doibase 10.1126/science.aaz8727}
  {\bibfield  {journal} {\bibinfo  {journal} {Science}\ }\textbf {\bibinfo
  {volume} {368}},\ \bibinfo {pages} {311} (\bibinfo {year}
  {2020})}\BibitemShut {NoStop}%
\bibitem [{\citenamefont {Zhang}\ \emph {et~al.}(2021)\citenamefont {Zhang},
  \citenamefont {Ouyang}, \citenamefont {Huang}, \citenamefont {Wang},
  \citenamefont {Zhang}, \citenamefont {Yu}, \citenamefont {Chang},
  \citenamefont {Liu}, \citenamefont {Deng},\ and\ \citenamefont
  {Duan}}]{PhysRevLett.127.090501}%
  \BibitemOpen
  \bibfield  {author} {\bibinfo {author} {\bibfnamefont {W.}~\bibnamefont
  {Zhang}}, \bibinfo {author} {\bibfnamefont {X.}~\bibnamefont {Ouyang}},
  \bibinfo {author} {\bibfnamefont {X.}~\bibnamefont {Huang}}, \bibinfo
  {author} {\bibfnamefont {X.}~\bibnamefont {Wang}}, \bibinfo {author}
  {\bibfnamefont {H.}~\bibnamefont {Zhang}}, \bibinfo {author} {\bibfnamefont
  {Y.}~\bibnamefont {Yu}}, \bibinfo {author} {\bibfnamefont {X.}~\bibnamefont
  {Chang}}, \bibinfo {author} {\bibfnamefont {Y.}~\bibnamefont {Liu}}, \bibinfo
  {author} {\bibfnamefont {D.-L.}\ \bibnamefont {Deng}}, \ and\ \bibinfo
  {author} {\bibfnamefont {L.-M.}\ \bibnamefont {Duan}},\ }\href {\doibase
  10.1103/PhysRevLett.127.090501} {\bibfield  {journal} {\bibinfo  {journal}
  {Phys. Rev. Lett.}\ }\textbf {\bibinfo {volume} {127}},\ \bibinfo {pages}
  {090501} (\bibinfo {year} {2021})}\BibitemShut {NoStop}%
\bibitem [{\citenamefont {Hodaei}\ \emph {et~al.}(2014)\citenamefont {Hodaei},
  \citenamefont {Miri}, \citenamefont {Heinrich}, \citenamefont
  {Christodoulides},\ and\ \citenamefont {Khajavikhan}}]{Hodaei975}%
  \BibitemOpen
  \bibfield  {author} {\bibinfo {author} {\bibfnamefont {H.}~\bibnamefont
  {Hodaei}}, \bibinfo {author} {\bibfnamefont {M.-A.}\ \bibnamefont {Miri}},
  \bibinfo {author} {\bibfnamefont {M.}~\bibnamefont {Heinrich}}, \bibinfo
  {author} {\bibfnamefont {D.~N.}\ \bibnamefont {Christodoulides}}, \ and\
  \bibinfo {author} {\bibfnamefont {M.}~\bibnamefont {Khajavikhan}},\ }\href
  {\doibase 10.1126/science.1258480} {\bibfield  {journal} {\bibinfo  {journal}
  {Science}\ }\textbf {\bibinfo {volume} {346}},\ \bibinfo {pages} {975}
  (\bibinfo {year} {2014})}\BibitemShut {NoStop}%
\bibitem [{\citenamefont {Feng}\ \emph {et~al.}(2014)\citenamefont {Feng},
  \citenamefont {Wong}, \citenamefont {Ma}, \citenamefont {Wang},\ and\
  \citenamefont {Zhang}}]{Feng972}%
  \BibitemOpen
  \bibfield  {author} {\bibinfo {author} {\bibfnamefont {L.}~\bibnamefont
  {Feng}}, \bibinfo {author} {\bibfnamefont {Z.~J.}\ \bibnamefont {Wong}},
  \bibinfo {author} {\bibfnamefont {R.-M.}\ \bibnamefont {Ma}}, \bibinfo
  {author} {\bibfnamefont {Y.}~\bibnamefont {Wang}}, \ and\ \bibinfo {author}
  {\bibfnamefont {X.}~\bibnamefont {Zhang}},\ }\href {\doibase
  10.1126/science.1258479} {\bibfield  {journal} {\bibinfo  {journal}
  {Science}\ }\textbf {\bibinfo {volume} {346}},\ \bibinfo {pages} {972}
  (\bibinfo {year} {2014})}\BibitemShut {NoStop}%
\bibitem [{\citenamefont {Harari}\ \emph {et~al.}(2018)\citenamefont {Harari},
  \citenamefont {Bandres}, \citenamefont {Lumer}, \citenamefont {Rechtsman},
  \citenamefont {Chong}, \citenamefont {Khajavikhan}, \citenamefont
  {Christodoulides},\ and\ \citenamefont {Segev}}]{Hararieaar4003}%
  \BibitemOpen
  \bibfield  {author} {\bibinfo {author} {\bibfnamefont {G.}~\bibnamefont
  {Harari}}, \bibinfo {author} {\bibfnamefont {M.~A.}\ \bibnamefont {Bandres}},
  \bibinfo {author} {\bibfnamefont {Y.}~\bibnamefont {Lumer}}, \bibinfo
  {author} {\bibfnamefont {M.~C.}\ \bibnamefont {Rechtsman}}, \bibinfo {author}
  {\bibfnamefont {Y.~D.}\ \bibnamefont {Chong}}, \bibinfo {author}
  {\bibfnamefont {M.}~\bibnamefont {Khajavikhan}}, \bibinfo {author}
  {\bibfnamefont {D.~N.}\ \bibnamefont {Christodoulides}}, \ and\ \bibinfo
  {author} {\bibfnamefont {M.}~\bibnamefont {Segev}},\ }\href
  {https://science.sciencemag.org/content/359/6381/eaar4003} {\bibfield
  {journal} {\bibinfo  {journal} {Science}\ }\textbf {\bibinfo {volume}
  {359}},\ \bibinfo {pages} {eaar4003} (\bibinfo {year} {2018})}\BibitemShut
  {NoStop}%
\bibitem [{\citenamefont {Regensburger}\ \emph {et~al.}(2012)\citenamefont
  {Regensburger}, \citenamefont {Bersch}, \citenamefont {Miri}, \citenamefont
  {Onishchukov}, \citenamefont {Christodoulides},\ and\ \citenamefont
  {Peschel}}]{Regensburger2012}%
  \BibitemOpen
  \bibfield  {author} {\bibinfo {author} {\bibfnamefont {A.}~\bibnamefont
  {Regensburger}}, \bibinfo {author} {\bibfnamefont {C.}~\bibnamefont
  {Bersch}}, \bibinfo {author} {\bibfnamefont {M.-A.}\ \bibnamefont {Miri}},
  \bibinfo {author} {\bibfnamefont {G.}~\bibnamefont {Onishchukov}}, \bibinfo
  {author} {\bibfnamefont {D.~N.}\ \bibnamefont {Christodoulides}}, \ and\
  \bibinfo {author} {\bibfnamefont {U.}~\bibnamefont {Peschel}},\ }\href
  {https://doi.org/10.1038/nature11298} {\bibfield  {journal} {\bibinfo
  {journal} {Nature (London)}\ }\textbf {\bibinfo {volume} {488}},\ \bibinfo
  {pages} {167} (\bibinfo {year} {2012})}\BibitemShut {NoStop}%
\bibitem [{\citenamefont {Hodaei}\ \emph {et~al.}(2017)\citenamefont {Hodaei},
  \citenamefont {Hassan}, \citenamefont {Wittek}, \citenamefont
  {Garcia-Gracia}, \citenamefont {El-Ganainy}, \citenamefont
  {Christodoulides},\ and\ \citenamefont {Khajavikhan}}]{EP2017}%
  \BibitemOpen
  \bibfield  {author} {\bibinfo {author} {\bibfnamefont {H.}~\bibnamefont
  {Hodaei}}, \bibinfo {author} {\bibfnamefont {A.~U.}\ \bibnamefont {Hassan}},
  \bibinfo {author} {\bibfnamefont {S.}~\bibnamefont {Wittek}}, \bibinfo
  {author} {\bibfnamefont {H.}~\bibnamefont {Garcia-Gracia}}, \bibinfo {author}
  {\bibfnamefont {R.}~\bibnamefont {El-Ganainy}}, \bibinfo {author}
  {\bibfnamefont {D.~N.}\ \bibnamefont {Christodoulides}}, \ and\ \bibinfo
  {author} {\bibfnamefont {M.}~\bibnamefont {Khajavikhan}},\ }\href
  {https://doi.org/10.1038/nature23280} {\bibfield  {journal} {\bibinfo
  {journal} {Nature (London)}\ }\textbf {\bibinfo {volume} {548}},\ \bibinfo
  {pages} {187} (\bibinfo {year} {2017})}\BibitemShut {NoStop}%
\bibitem [{\citenamefont {Chen}\ \emph
  {et~al.}(2017{\natexlab{a}})\citenamefont {Chen}, \citenamefont
  {Kaya~\"{O}zdemir}, \citenamefont {Zhao}, \citenamefont {Wiersig},\ and\
  \citenamefont {Yang}}]{Chen2017}%
  \BibitemOpen
  \bibfield  {author} {\bibinfo {author} {\bibfnamefont {W.}~\bibnamefont
  {Chen}}, \bibinfo {author} {\bibfnamefont {c.}~\bibnamefont
  {Kaya~\"{O}zdemir}}, \bibinfo {author} {\bibfnamefont {G.}~\bibnamefont
  {Zhao}}, \bibinfo {author} {\bibfnamefont {J.}~\bibnamefont {Wiersig}}, \
  and\ \bibinfo {author} {\bibfnamefont {L.}~\bibnamefont {Yang}},\ }\href
  {https://doi.org/10.1038/nature23281} {\bibfield  {journal} {\bibinfo
  {journal} {Nature (London)}\ }\textbf {\bibinfo {volume} {548}},\ \bibinfo
  {pages} {192} (\bibinfo {year} {2017}{\natexlab{a}})}\BibitemShut {NoStop}%
\bibitem [{\citenamefont {Armitage}\ \emph {et~al.}(2018)\citenamefont
  {Armitage}, \citenamefont {Mele},\ and\ \citenamefont
  {Vishwanath}}]{RevModPhys.90.015001}%
  \BibitemOpen
  \bibfield  {author} {\bibinfo {author} {\bibfnamefont {N.~P.}\ \bibnamefont
  {Armitage}}, \bibinfo {author} {\bibfnamefont {E.~J.}\ \bibnamefont {Mele}},
  \ and\ \bibinfo {author} {\bibfnamefont {A.}~\bibnamefont {Vishwanath}},\
  }\href {\doibase 10.1103/RevModPhys.90.015001} {\bibfield  {journal}
  {\bibinfo  {journal} {Rev. Mod. Phys.}\ }\textbf {\bibinfo {volume} {90}},\
  \bibinfo {pages} {015001} (\bibinfo {year} {2018})}\BibitemShut {NoStop}%
\bibitem [{\citenamefont {Burkov}\ \emph {et~al.}(2011)\citenamefont {Burkov},
  \citenamefont {Hook},\ and\ \citenamefont {Balents}}]{PhysRevB.84.235126}%
  \BibitemOpen
  \bibfield  {author} {\bibinfo {author} {\bibfnamefont {A.~A.}\ \bibnamefont
  {Burkov}}, \bibinfo {author} {\bibfnamefont {M.~D.}\ \bibnamefont {Hook}}, \
  and\ \bibinfo {author} {\bibfnamefont {L.}~\bibnamefont {Balents}},\ }\href
  {\doibase 10.1103/PhysRevB.84.235126} {\bibfield  {journal} {\bibinfo
  {journal} {Phys. Rev. B}\ }\textbf {\bibinfo {volume} {84}},\ \bibinfo
  {pages} {235126} (\bibinfo {year} {2011})}\BibitemShut {NoStop}%
\bibitem [{\citenamefont {Wang}\ \emph {et~al.}(2020)\citenamefont {Wang},
  \citenamefont {Lin}, \citenamefont {Jiang}, \citenamefont {Guo},\ and\
  \citenamefont {Jiang}}]{PhysRevLett.125.146401}%
  \BibitemOpen
  \bibfield  {author} {\bibinfo {author} {\bibfnamefont {H.-X.}\ \bibnamefont
  {Wang}}, \bibinfo {author} {\bibfnamefont {Z.-K.}\ \bibnamefont {Lin}},
  \bibinfo {author} {\bibfnamefont {B.}~\bibnamefont {Jiang}}, \bibinfo
  {author} {\bibfnamefont {G.-Y.}\ \bibnamefont {Guo}}, \ and\ \bibinfo
  {author} {\bibfnamefont {J.-H.}\ \bibnamefont {Jiang}},\ }\href {\doibase
  10.1103/PhysRevLett.125.146401} {\bibfield  {journal} {\bibinfo  {journal}
  {Phys. Rev. Lett.}\ }\textbf {\bibinfo {volume} {125}},\ \bibinfo {pages}
  {146401} (\bibinfo {year} {2020})}\BibitemShut {NoStop}%
\bibitem [{\citenamefont {Xu}\ \emph {et~al.}(2015)\citenamefont {Xu},
  \citenamefont {Belopolski}, \citenamefont {Alidoust}, \citenamefont
  {Neupane}, \citenamefont {Bian}, \citenamefont {Zhang}, \citenamefont
  {Sankar}, \citenamefont {Chang}, \citenamefont {Yuan}, \citenamefont {Lee},
  \citenamefont {Huang}, \citenamefont {Zheng}, \citenamefont {Ma},
  \citenamefont {Sanchez}, \citenamefont {Wang}, \citenamefont {Bansil},
  \citenamefont {Chou}, \citenamefont {Shibayev}, \citenamefont {Lin},
  \citenamefont {Jia},\ and\ \citenamefont {Hasan}}]{Xu2015}%
  \BibitemOpen
  \bibfield  {author} {\bibinfo {author} {\bibfnamefont {S.-Y.}\ \bibnamefont
  {Xu}}, \bibinfo {author} {\bibfnamefont {I.}~\bibnamefont {Belopolski}},
  \bibinfo {author} {\bibfnamefont {N.}~\bibnamefont {Alidoust}}, \bibinfo
  {author} {\bibfnamefont {M.}~\bibnamefont {Neupane}}, \bibinfo {author}
  {\bibfnamefont {G.}~\bibnamefont {Bian}}, \bibinfo {author} {\bibfnamefont
  {C.}~\bibnamefont {Zhang}}, \bibinfo {author} {\bibfnamefont
  {R.}~\bibnamefont {Sankar}}, \bibinfo {author} {\bibfnamefont
  {G.}~\bibnamefont {Chang}}, \bibinfo {author} {\bibfnamefont
  {Z.}~\bibnamefont {Yuan}}, \bibinfo {author} {\bibfnamefont {C.-C.}\
  \bibnamefont {Lee}}, \bibinfo {author} {\bibfnamefont {S.-M.}\ \bibnamefont
  {Huang}}, \bibinfo {author} {\bibfnamefont {H.}~\bibnamefont {Zheng}},
  \bibinfo {author} {\bibfnamefont {J.}~\bibnamefont {Ma}}, \bibinfo {author}
  {\bibfnamefont {D.~S.}\ \bibnamefont {Sanchez}}, \bibinfo {author}
  {\bibfnamefont {B.}~\bibnamefont {Wang}}, \bibinfo {author} {\bibfnamefont
  {A.}~\bibnamefont {Bansil}}, \bibinfo {author} {\bibfnamefont
  {F.}~\bibnamefont {Chou}}, \bibinfo {author} {\bibfnamefont {P.~P.}\
  \bibnamefont {Shibayev}}, \bibinfo {author} {\bibfnamefont {H.}~\bibnamefont
  {Lin}}, \bibinfo {author} {\bibfnamefont {S.}~\bibnamefont {Jia}}, \ and\
  \bibinfo {author} {\bibfnamefont {M.~Z.}\ \bibnamefont {Hasan}},\ }\href
  {\doibase 10.1126/science.aaa9297} {\bibfield  {journal} {\bibinfo  {journal}
  {Science}\ }\textbf {\bibinfo {volume} {349}},\ \bibinfo {pages} {613}
  (\bibinfo {year} {2015})}\BibitemShut {NoStop}%
\bibitem [{\citenamefont {Lv}\ \emph {et~al.}(2015)\citenamefont {Lv},
  \citenamefont {Weng}, \citenamefont {Fu}, \citenamefont {Wang}, \citenamefont
  {Miao}, \citenamefont {Ma}, \citenamefont {Richard}, \citenamefont {Huang},
  \citenamefont {Zhao}, \citenamefont {Chen}, \citenamefont {Fang},
  \citenamefont {Dai}, \citenamefont {Qian},\ and\ \citenamefont
  {Ding}}]{PhysRevX.5.031013}%
  \BibitemOpen
  \bibfield  {author} {\bibinfo {author} {\bibfnamefont {B.~Q.}\ \bibnamefont
  {Lv}}, \bibinfo {author} {\bibfnamefont {H.~M.}\ \bibnamefont {Weng}},
  \bibinfo {author} {\bibfnamefont {B.~B.}\ \bibnamefont {Fu}}, \bibinfo
  {author} {\bibfnamefont {X.~P.}\ \bibnamefont {Wang}}, \bibinfo {author}
  {\bibfnamefont {H.}~\bibnamefont {Miao}}, \bibinfo {author} {\bibfnamefont
  {J.}~\bibnamefont {Ma}}, \bibinfo {author} {\bibfnamefont {P.}~\bibnamefont
  {Richard}}, \bibinfo {author} {\bibfnamefont {X.~C.}\ \bibnamefont {Huang}},
  \bibinfo {author} {\bibfnamefont {L.~X.}\ \bibnamefont {Zhao}}, \bibinfo
  {author} {\bibfnamefont {G.~F.}\ \bibnamefont {Chen}}, \bibinfo {author}
  {\bibfnamefont {Z.}~\bibnamefont {Fang}}, \bibinfo {author} {\bibfnamefont
  {X.}~\bibnamefont {Dai}}, \bibinfo {author} {\bibfnamefont {T.}~\bibnamefont
  {Qian}}, \ and\ \bibinfo {author} {\bibfnamefont {H.}~\bibnamefont {Ding}},\
  }\href {\doibase 10.1103/PhysRevX.5.031013} {\bibfield  {journal} {\bibinfo
  {journal} {Phys. Rev. X}\ }\textbf {\bibinfo {volume} {5}},\ \bibinfo {pages}
  {031013} (\bibinfo {year} {2015})}\BibitemShut {NoStop}%
\bibitem [{\citenamefont {Li}\ \emph {et~al.}(2018{\natexlab{a}})\citenamefont
  {Li}, \citenamefont {Wang}, \citenamefont {Wan}, \citenamefont {Wan},
  \citenamefont {Lu},\ and\ \citenamefont {Xie}}]{PhysRevLett.120.146602}%
  \BibitemOpen
  \bibfield  {author} {\bibinfo {author} {\bibfnamefont {C.}~\bibnamefont
  {Li}}, \bibinfo {author} {\bibfnamefont {C.~M.}\ \bibnamefont {Wang}},
  \bibinfo {author} {\bibfnamefont {B.}~\bibnamefont {Wan}}, \bibinfo {author}
  {\bibfnamefont {X.}~\bibnamefont {Wan}}, \bibinfo {author} {\bibfnamefont
  {H.-Z.}\ \bibnamefont {Lu}}, \ and\ \bibinfo {author} {\bibfnamefont {X.~C.}\
  \bibnamefont {Xie}},\ }\href {\doibase 10.1103/PhysRevLett.120.146602}
  {\bibfield  {journal} {\bibinfo  {journal} {Phys. Rev. Lett.}\ }\textbf
  {\bibinfo {volume} {120}},\ \bibinfo {pages} {146602} (\bibinfo {year}
  {2018}{\natexlab{a}})}\BibitemShut {NoStop}%
\bibitem [{\citenamefont {Ahn}\ \emph {et~al.}(2018)\citenamefont {Ahn},
  \citenamefont {Kim}, \citenamefont {Kim},\ and\ \citenamefont
  {Yang}}]{PhysRevLett.121.106403}%
  \BibitemOpen
  \bibfield  {author} {\bibinfo {author} {\bibfnamefont {J.}~\bibnamefont
  {Ahn}}, \bibinfo {author} {\bibfnamefont {D.}~\bibnamefont {Kim}}, \bibinfo
  {author} {\bibfnamefont {Y.}~\bibnamefont {Kim}}, \ and\ \bibinfo {author}
  {\bibfnamefont {B.-J.}\ \bibnamefont {Yang}},\ }\href {\doibase
  10.1103/PhysRevLett.121.106403} {\bibfield  {journal} {\bibinfo  {journal}
  {Phys. Rev. Lett.}\ }\textbf {\bibinfo {volume} {121}},\ \bibinfo {pages}
  {106403} (\bibinfo {year} {2018})}\BibitemShut {NoStop}%
\bibitem [{\citenamefont {Wang}\ \emph {et~al.}(2018)\citenamefont {Wang},
  \citenamefont {Nie}, \citenamefont {Weng}, \citenamefont {Kawazoe},\ and\
  \citenamefont {Chen}}]{PhysRevLett.120.026402}%
  \BibitemOpen
  \bibfield  {author} {\bibinfo {author} {\bibfnamefont {J.-T.}\ \bibnamefont
  {Wang}}, \bibinfo {author} {\bibfnamefont {S.}~\bibnamefont {Nie}}, \bibinfo
  {author} {\bibfnamefont {H.}~\bibnamefont {Weng}}, \bibinfo {author}
  {\bibfnamefont {Y.}~\bibnamefont {Kawazoe}}, \ and\ \bibinfo {author}
  {\bibfnamefont {C.}~\bibnamefont {Chen}},\ }\href {\doibase
  10.1103/PhysRevLett.120.026402} {\bibfield  {journal} {\bibinfo  {journal}
  {Phys. Rev. Lett.}\ }\textbf {\bibinfo {volume} {120}},\ \bibinfo {pages}
  {026402} (\bibinfo {year} {2018})}\BibitemShut {NoStop}%
\bibitem [{\citenamefont {Chen}\ \emph
  {et~al.}(2017{\natexlab{b}})\citenamefont {Chen}, \citenamefont {Lu},\ and\
  \citenamefont {Hou}}]{PhysRevB.96.041102}%
  \BibitemOpen
  \bibfield  {author} {\bibinfo {author} {\bibfnamefont {W.}~\bibnamefont
  {Chen}}, \bibinfo {author} {\bibfnamefont {H.-Z.}\ \bibnamefont {Lu}}, \ and\
  \bibinfo {author} {\bibfnamefont {J.-M.}\ \bibnamefont {Hou}},\ }\href
  {\doibase 10.1103/PhysRevB.96.041102} {\bibfield  {journal} {\bibinfo
  {journal} {Phys. Rev. B}\ }\textbf {\bibinfo {volume} {96}},\ \bibinfo
  {pages} {041102(R)} (\bibinfo {year} {2017}{\natexlab{b}})}\BibitemShut
  {NoStop}%
\bibitem [{\citenamefont {Zhou}\ \emph {et~al.}(2018)\citenamefont {Zhou},
  \citenamefont {Xiong}, \citenamefont {Wan},\ and\ \citenamefont
  {An}}]{PhysRevB.97.155140}%
  \BibitemOpen
  \bibfield  {author} {\bibinfo {author} {\bibfnamefont {Y.}~\bibnamefont
  {Zhou}}, \bibinfo {author} {\bibfnamefont {F.}~\bibnamefont {Xiong}},
  \bibinfo {author} {\bibfnamefont {X.}~\bibnamefont {Wan}}, \ and\ \bibinfo
  {author} {\bibfnamefont {J.}~\bibnamefont {An}},\ }\href {\doibase
  10.1103/PhysRevB.97.155140} {\bibfield  {journal} {\bibinfo  {journal} {Phys.
  Rev. B}\ }\textbf {\bibinfo {volume} {97}},\ \bibinfo {pages} {155140}
  (\bibinfo {year} {2018})}\BibitemShut {NoStop}%
\bibitem [{\citenamefont {Ezawa}(2016)}]{PhysRevLett.116.127202}%
  \BibitemOpen
  \bibfield  {author} {\bibinfo {author} {\bibfnamefont {M.}~\bibnamefont
  {Ezawa}},\ }\href {\doibase 10.1103/PhysRevLett.116.127202} {\bibfield
  {journal} {\bibinfo  {journal} {Phys. Rev. Lett.}\ }\textbf {\bibinfo
  {volume} {116}},\ \bibinfo {pages} {127202} (\bibinfo {year}
  {2016})}\BibitemShut {NoStop}%
\bibitem [{\citenamefont {Cuong}\ \emph {et~al.}(2020)\citenamefont {Cuong},
  \citenamefont {Tateishi}, \citenamefont {Cameau}, \citenamefont {Niibe},
  \citenamefont {Umezawa}, \citenamefont {Slater}, \citenamefont {Yubuta},
  \citenamefont {Kondo}, \citenamefont {Ogata}, \citenamefont {Okada},\ and\
  \citenamefont {Matsuda}}]{PhysRevB.101.195412}%
  \BibitemOpen
  \bibfield  {author} {\bibinfo {author} {\bibfnamefont {N.~T.}\ \bibnamefont
  {Cuong}}, \bibinfo {author} {\bibfnamefont {I.}~\bibnamefont {Tateishi}},
  \bibinfo {author} {\bibfnamefont {M.}~\bibnamefont {Cameau}}, \bibinfo
  {author} {\bibfnamefont {M.}~\bibnamefont {Niibe}}, \bibinfo {author}
  {\bibfnamefont {N.}~\bibnamefont {Umezawa}}, \bibinfo {author} {\bibfnamefont
  {B.}~\bibnamefont {Slater}}, \bibinfo {author} {\bibfnamefont
  {K.}~\bibnamefont {Yubuta}}, \bibinfo {author} {\bibfnamefont
  {T.}~\bibnamefont {Kondo}}, \bibinfo {author} {\bibfnamefont
  {M.}~\bibnamefont {Ogata}}, \bibinfo {author} {\bibfnamefont
  {S.}~\bibnamefont {Okada}}, \ and\ \bibinfo {author} {\bibfnamefont
  {I.}~\bibnamefont {Matsuda}},\ }\href {\doibase 10.1103/PhysRevB.101.195412}
  {\bibfield  {journal} {\bibinfo  {journal} {Phys. Rev. B}\ }\textbf {\bibinfo
  {volume} {101}},\ \bibinfo {pages} {195412} (\bibinfo {year}
  {2020})}\BibitemShut {NoStop}%
\bibitem [{\citenamefont {Zheng}\ \emph {et~al.}(2017)\citenamefont {Zheng},
  \citenamefont {Chang}, \citenamefont {Huang}, \citenamefont {Guo},
  \citenamefont {Zhang}, \citenamefont {Zhang}, \citenamefont {Yin},
  \citenamefont {Xu}, \citenamefont {Belopolski}, \citenamefont {Alidoust},
  \citenamefont {Sanchez}, \citenamefont {Bian}, \citenamefont {Chang},
  \citenamefont {Neupert}, \citenamefont {Jeng}, \citenamefont {Jia},
  \citenamefont {Lin},\ and\ \citenamefont {Hasan}}]{PhysRevLett.119.196403}%
  \BibitemOpen
  \bibfield  {author} {\bibinfo {author} {\bibfnamefont {H.}~\bibnamefont
  {Zheng}}, \bibinfo {author} {\bibfnamefont {G.}~\bibnamefont {Chang}},
  \bibinfo {author} {\bibfnamefont {S.-M.}\ \bibnamefont {Huang}}, \bibinfo
  {author} {\bibfnamefont {C.}~\bibnamefont {Guo}}, \bibinfo {author}
  {\bibfnamefont {X.}~\bibnamefont {Zhang}}, \bibinfo {author} {\bibfnamefont
  {S.}~\bibnamefont {Zhang}}, \bibinfo {author} {\bibfnamefont
  {J.}~\bibnamefont {Yin}}, \bibinfo {author} {\bibfnamefont {S.-Y.}\
  \bibnamefont {Xu}}, \bibinfo {author} {\bibfnamefont {I.}~\bibnamefont
  {Belopolski}}, \bibinfo {author} {\bibfnamefont {N.}~\bibnamefont
  {Alidoust}}, \bibinfo {author} {\bibfnamefont {D.~S.}\ \bibnamefont
  {Sanchez}}, \bibinfo {author} {\bibfnamefont {G.}~\bibnamefont {Bian}},
  \bibinfo {author} {\bibfnamefont {T.-R.}\ \bibnamefont {Chang}}, \bibinfo
  {author} {\bibfnamefont {T.}~\bibnamefont {Neupert}}, \bibinfo {author}
  {\bibfnamefont {H.-T.}\ \bibnamefont {Jeng}}, \bibinfo {author}
  {\bibfnamefont {S.}~\bibnamefont {Jia}}, \bibinfo {author} {\bibfnamefont
  {H.}~\bibnamefont {Lin}}, \ and\ \bibinfo {author} {\bibfnamefont {M.~Z.}\
  \bibnamefont {Hasan}},\ }\href {\doibase 10.1103/PhysRevLett.119.196403}
  {\bibfield  {journal} {\bibinfo  {journal} {Phys. Rev. Lett.}\ }\textbf
  {\bibinfo {volume} {119}},\ \bibinfo {pages} {196403} (\bibinfo {year}
  {2017})}\BibitemShut {NoStop}%
\bibitem [{\citenamefont {Zyuzin}\ and\ \citenamefont
  {Zyuzin}(2018)}]{PhysRevB.97.041203}%
  \BibitemOpen
  \bibfield  {author} {\bibinfo {author} {\bibfnamefont {A.~A.}\ \bibnamefont
  {Zyuzin}}\ and\ \bibinfo {author} {\bibfnamefont {A.~Y.}\ \bibnamefont
  {Zyuzin}},\ }\href {\doibase 10.1103/PhysRevB.97.041203} {\bibfield
  {journal} {\bibinfo  {journal} {Phys. Rev. B}\ }\textbf {\bibinfo {volume}
  {97}},\ \bibinfo {pages} {041203(R)} (\bibinfo {year} {2018})}\BibitemShut
  {NoStop}%
\bibitem [{\citenamefont {Budich}\ \emph {et~al.}(2019)\citenamefont {Budich},
  \citenamefont {Carlstr\"om}, \citenamefont {Kunst},\ and\ \citenamefont
  {Bergholtz}}]{PhysRevB.99.041406}%
  \BibitemOpen
  \bibfield  {author} {\bibinfo {author} {\bibfnamefont {J.~C.}\ \bibnamefont
  {Budich}}, \bibinfo {author} {\bibfnamefont {J.}~\bibnamefont {Carlstr\"om}},
  \bibinfo {author} {\bibfnamefont {F.~K.}\ \bibnamefont {Kunst}}, \ and\
  \bibinfo {author} {\bibfnamefont {E.~J.}\ \bibnamefont {Bergholtz}},\ }\href
  {\doibase 10.1103/PhysRevB.99.041406} {\bibfield  {journal} {\bibinfo
  {journal} {Phys. Rev. B}\ }\textbf {\bibinfo {volume} {99}},\ \bibinfo
  {pages} {041406(R)} (\bibinfo {year} {2019})}\BibitemShut {NoStop}%
\bibitem [{\citenamefont {Wang}\ \emph {et~al.}(2019)\citenamefont {Wang},
  \citenamefont {Ruan},\ and\ \citenamefont {Zhang}}]{PhysRevB.99.075130}%
  \BibitemOpen
  \bibfield  {author} {\bibinfo {author} {\bibfnamefont {H.}~\bibnamefont
  {Wang}}, \bibinfo {author} {\bibfnamefont {J.}~\bibnamefont {Ruan}}, \ and\
  \bibinfo {author} {\bibfnamefont {H.}~\bibnamefont {Zhang}},\ }\href
  {\doibase 10.1103/PhysRevB.99.075130} {\bibfield  {journal} {\bibinfo
  {journal} {Phys. Rev. B}\ }\textbf {\bibinfo {volume} {99}},\ \bibinfo
  {pages} {075130} (\bibinfo {year} {2019})}\BibitemShut {NoStop}%
\bibitem [{\citenamefont {Rui}\ \emph {et~al.}(2019)\citenamefont {Rui},
  \citenamefont {Hirschmann},\ and\ \citenamefont
  {Schnyder}}]{PhysRevB.100.245116}%
  \BibitemOpen
  \bibfield  {author} {\bibinfo {author} {\bibfnamefont {W.~B.}\ \bibnamefont
  {Rui}}, \bibinfo {author} {\bibfnamefont {M.~M.}\ \bibnamefont {Hirschmann}},
  \ and\ \bibinfo {author} {\bibfnamefont {A.~P.}\ \bibnamefont {Schnyder}},\
  }\href {\doibase 10.1103/PhysRevB.100.245116} {\bibfield  {journal} {\bibinfo
   {journal} {Phys. Rev. B}\ }\textbf {\bibinfo {volume} {100}},\ \bibinfo
  {pages} {245116} (\bibinfo {year} {2019})}\BibitemShut {NoStop}%
\bibitem [{\citenamefont {Zhang}\ \emph
  {et~al.}(2020{\natexlab{b}})\citenamefont {Zhang}, \citenamefont {Yang},\
  and\ \citenamefont {Hu}}]{PhysRevB.102.045412}%
  \BibitemOpen
  \bibfield  {author} {\bibinfo {author} {\bibfnamefont {Z.}~\bibnamefont
  {Zhang}}, \bibinfo {author} {\bibfnamefont {Z.}~\bibnamefont {Yang}}, \ and\
  \bibinfo {author} {\bibfnamefont {J.}~\bibnamefont {Hu}},\ }\href {\doibase
  10.1103/PhysRevB.102.045412} {\bibfield  {journal} {\bibinfo  {journal}
  {Phys. Rev. B}\ }\textbf {\bibinfo {volume} {102}},\ \bibinfo {pages}
  {045412} (\bibinfo {year} {2020}{\natexlab{b}})}\BibitemShut {NoStop}%
\bibitem [{\citenamefont {Molina}\ and\ \citenamefont
  {Gonz\'alez}(2018)}]{PhysRevLett.120.146601}%
  \BibitemOpen
  \bibfield  {author} {\bibinfo {author} {\bibfnamefont {R.~A.}\ \bibnamefont
  {Molina}}\ and\ \bibinfo {author} {\bibfnamefont {J.}~\bibnamefont
  {Gonz\'alez}},\ }\href {\doibase 10.1103/PhysRevLett.120.146601} {\bibfield
  {journal} {\bibinfo  {journal} {Phys. Rev. Lett.}\ }\textbf {\bibinfo
  {volume} {120}},\ \bibinfo {pages} {146601} (\bibinfo {year}
  {2018})}\BibitemShut {NoStop}%
\bibitem [{\citenamefont {Xue}\ \emph {et~al.}(2020)\citenamefont {Xue},
  \citenamefont {Wang}, \citenamefont {Zhang},\ and\ \citenamefont
  {Chong}}]{PhysRevLett.124.236403}%
  \BibitemOpen
  \bibfield  {author} {\bibinfo {author} {\bibfnamefont {H.}~\bibnamefont
  {Xue}}, \bibinfo {author} {\bibfnamefont {Q.}~\bibnamefont {Wang}}, \bibinfo
  {author} {\bibfnamefont {B.}~\bibnamefont {Zhang}}, \ and\ \bibinfo {author}
  {\bibfnamefont {Y.~D.}\ \bibnamefont {Chong}},\ }\href {\doibase
  10.1103/PhysRevLett.124.236403} {\bibfield  {journal} {\bibinfo  {journal}
  {Phys. Rev. Lett.}\ }\textbf {\bibinfo {volume} {124}},\ \bibinfo {pages}
  {236403} (\bibinfo {year} {2020})}\BibitemShut {NoStop}%
\bibitem [{\citenamefont {Hu}\ and\ \citenamefont
  {Zhao}(2021)}]{PhysRevLett.126.010401}%
  \BibitemOpen
  \bibfield  {author} {\bibinfo {author} {\bibfnamefont {H.}~\bibnamefont
  {Hu}}\ and\ \bibinfo {author} {\bibfnamefont {E.}~\bibnamefont {Zhao}},\
  }\href {\doibase 10.1103/PhysRevLett.126.010401} {\bibfield  {journal}
  {\bibinfo  {journal} {Phys. Rev. Lett.}\ }\textbf {\bibinfo {volume} {126}},\
  \bibinfo {pages} {010401} (\bibinfo {year} {2021})}\BibitemShut {NoStop}%
\bibitem [{\citenamefont {Ghorashi}\ \emph {et~al.}(2021)\citenamefont
  {Ghorashi}, \citenamefont {Li},\ and\ \citenamefont
  {Sato}}]{PhysRevB.104.L161117}%
  \BibitemOpen
  \bibfield  {author} {\bibinfo {author} {\bibfnamefont {S.~A.~A.}\
  \bibnamefont {Ghorashi}}, \bibinfo {author} {\bibfnamefont {T.}~\bibnamefont
  {Li}}, \ and\ \bibinfo {author} {\bibfnamefont {M.}~\bibnamefont {Sato}},\
  }\href {\doibase 10.1103/PhysRevB.104.L161117} {\bibfield  {journal}
  {\bibinfo  {journal} {Phys. Rev. B}\ }\textbf {\bibinfo {volume} {104}},\
  \bibinfo {pages} {L161117} (\bibinfo {year} {2021})}\BibitemShut {NoStop}%
\bibitem [{\citenamefont {Tzeng}\ \emph {et~al.}(2021)\citenamefont {Tzeng},
  \citenamefont {Ju}, \citenamefont {Chen},\ and\ \citenamefont
  {Huang}}]{PhysRevResearch.3.013015}%
  \BibitemOpen
  \bibfield  {author} {\bibinfo {author} {\bibfnamefont {Y.-C.}\ \bibnamefont
  {Tzeng}}, \bibinfo {author} {\bibfnamefont {C.-Y.}\ \bibnamefont {Ju}},
  \bibinfo {author} {\bibfnamefont {G.-Y.}\ \bibnamefont {Chen}}, \ and\
  \bibinfo {author} {\bibfnamefont {W.-M.}\ \bibnamefont {Huang}},\ }\href
  {\doibase 10.1103/PhysRevResearch.3.013015} {\bibfield  {journal} {\bibinfo
  {journal} {Phys. Rev. Research}\ }\textbf {\bibinfo {volume} {3}},\ \bibinfo
  {pages} {013015} (\bibinfo {year} {2021})}\BibitemShut {NoStop}%
\bibitem [{\citenamefont {Kawabata}\ \emph
  {et~al.}(2019{\natexlab{b}})\citenamefont {Kawabata}, \citenamefont
  {Bessho},\ and\ \citenamefont {Sato}}]{PhysRevLett.123.066405}%
  \BibitemOpen
  \bibfield  {author} {\bibinfo {author} {\bibfnamefont {K.}~\bibnamefont
  {Kawabata}}, \bibinfo {author} {\bibfnamefont {T.}~\bibnamefont {Bessho}}, \
  and\ \bibinfo {author} {\bibfnamefont {M.}~\bibnamefont {Sato}},\ }\href
  {\doibase 10.1103/PhysRevLett.123.066405} {\bibfield  {journal} {\bibinfo
  {journal} {Phys. Rev. Lett.}\ }\textbf {\bibinfo {volume} {123}},\ \bibinfo
  {pages} {066405} (\bibinfo {year} {2019}{\natexlab{b}})}\BibitemShut
  {NoStop}%
\bibitem [{\citenamefont {Hou}\ \emph {et~al.}(2020)\citenamefont {Hou},
  \citenamefont {Li}, \citenamefont {Luo}, \citenamefont {Gu},\ and\
  \citenamefont {Zhang}}]{PhysRevLett.124.073603}%
  \BibitemOpen
  \bibfield  {author} {\bibinfo {author} {\bibfnamefont {J.}~\bibnamefont
  {Hou}}, \bibinfo {author} {\bibfnamefont {Z.}~\bibnamefont {Li}}, \bibinfo
  {author} {\bibfnamefont {X.-W.}\ \bibnamefont {Luo}}, \bibinfo {author}
  {\bibfnamefont {Q.}~\bibnamefont {Gu}}, \ and\ \bibinfo {author}
  {\bibfnamefont {C.}~\bibnamefont {Zhang}},\ }\href {\doibase
  10.1103/PhysRevLett.124.073603} {\bibfield  {journal} {\bibinfo  {journal}
  {Phys. Rev. Lett.}\ }\textbf {\bibinfo {volume} {124}},\ \bibinfo {pages}
  {073603} (\bibinfo {year} {2020})}\BibitemShut {NoStop}%
\bibitem [{\citenamefont {Cerjan}\ \emph {et~al.}(2018)\citenamefont {Cerjan},
  \citenamefont {Xiao}, \citenamefont {Yuan},\ and\ \citenamefont
  {Fan}}]{PhysRevB.97.075128}%
  \BibitemOpen
  \bibfield  {author} {\bibinfo {author} {\bibfnamefont {A.}~\bibnamefont
  {Cerjan}}, \bibinfo {author} {\bibfnamefont {M.}~\bibnamefont {Xiao}},
  \bibinfo {author} {\bibfnamefont {L.}~\bibnamefont {Yuan}}, \ and\ \bibinfo
  {author} {\bibfnamefont {S.}~\bibnamefont {Fan}},\ }\href {\doibase
  10.1103/PhysRevB.97.075128} {\bibfield  {journal} {\bibinfo  {journal} {Phys.
  Rev. B}\ }\textbf {\bibinfo {volume} {97}},\ \bibinfo {pages} {075128}
  (\bibinfo {year} {2018})}\BibitemShut {NoStop}%
\bibitem [{\citenamefont {Cerjan}\ \emph {et~al.}(2019)\citenamefont {Cerjan},
  \citenamefont {Huang}, \citenamefont {Wang}, \citenamefont {Chen},
  \citenamefont {Chong},\ and\ \citenamefont {Rechtsman}}]{Cerjan2019}%
  \BibitemOpen
  \bibfield  {author} {\bibinfo {author} {\bibfnamefont {A.}~\bibnamefont
  {Cerjan}}, \bibinfo {author} {\bibfnamefont {S.}~\bibnamefont {Huang}},
  \bibinfo {author} {\bibfnamefont {M.}~\bibnamefont {Wang}}, \bibinfo {author}
  {\bibfnamefont {K.~P.}\ \bibnamefont {Chen}}, \bibinfo {author}
  {\bibfnamefont {Y.}~\bibnamefont {Chong}}, \ and\ \bibinfo {author}
  {\bibfnamefont {M.~C.}\ \bibnamefont {Rechtsman}},\ }\href {\doibase
  10.1038/s41566-019-0453-z} {\bibfield  {journal} {\bibinfo  {journal} {Nature
  Photonics}\ }\textbf {\bibinfo {volume} {13}},\ \bibinfo {pages} {623}
  (\bibinfo {year} {2019})}\BibitemShut {NoStop}%
\bibitem [{\citenamefont {Matsushita}\ \emph {et~al.}(2019)\citenamefont
  {Matsushita}, \citenamefont {Nagai},\ and\ \citenamefont
  {Fujimoto}}]{PhysRevB.100.245205}%
  \BibitemOpen
  \bibfield  {author} {\bibinfo {author} {\bibfnamefont {T.}~\bibnamefont
  {Matsushita}}, \bibinfo {author} {\bibfnamefont {Y.}~\bibnamefont {Nagai}}, \
  and\ \bibinfo {author} {\bibfnamefont {S.}~\bibnamefont {Fujimoto}},\ }\href
  {\doibase 10.1103/PhysRevB.100.245205} {\bibfield  {journal} {\bibinfo
  {journal} {Phys. Rev. B}\ }\textbf {\bibinfo {volume} {100}},\ \bibinfo
  {pages} {245205} (\bibinfo {year} {2019})}\BibitemShut {NoStop}%
\bibitem [{\citenamefont {Mc~Guinness}\ and\ \citenamefont
  {Eastham}(2020)}]{PhysRevResearch.2.043268}%
  \BibitemOpen
  \bibfield  {author} {\bibinfo {author} {\bibfnamefont {R.~L.}\ \bibnamefont
  {Mc~Guinness}}\ and\ \bibinfo {author} {\bibfnamefont {P.~R.}\ \bibnamefont
  {Eastham}},\ }\href {\doibase 10.1103/PhysRevResearch.2.043268} {\bibfield
  {journal} {\bibinfo  {journal} {Phys. Rev. Research}\ }\textbf {\bibinfo
  {volume} {2}},\ \bibinfo {pages} {043268} (\bibinfo {year}
  {2020})}\BibitemShut {NoStop}%
\bibitem [{\citenamefont {Liu}\ \emph {et~al.}(2021)\citenamefont {Liu},
  \citenamefont {He}, \citenamefont {Yang},\ and\ \citenamefont
  {Nori}}]{PhysRevLett.127.196801}%
  \BibitemOpen
  \bibfield  {author} {\bibinfo {author} {\bibfnamefont {T.}~\bibnamefont
  {Liu}}, \bibinfo {author} {\bibfnamefont {J.~J.}\ \bibnamefont {He}},
  \bibinfo {author} {\bibfnamefont {Z.}~\bibnamefont {Yang}}, \ and\ \bibinfo
  {author} {\bibfnamefont {F.}~\bibnamefont {Nori}},\ }\href {\doibase
  10.1103/PhysRevLett.127.196801} {\bibfield  {journal} {\bibinfo  {journal}
  {Phys. Rev. Lett.}\ }\textbf {\bibinfo {volume} {127}},\ \bibinfo {pages}
  {196801} (\bibinfo {year} {2021})}\BibitemShut {NoStop}%
\bibitem [{\citenamefont {Chowdhury}\ \emph {et~al.}(2021)\citenamefont
  {Chowdhury}, \citenamefont {Banerjee},\ and\ \citenamefont
  {Narayan}}]{PhysRevA.103.L051101}%
  \BibitemOpen
  \bibfield  {author} {\bibinfo {author} {\bibfnamefont {D.}~\bibnamefont
  {Chowdhury}}, \bibinfo {author} {\bibfnamefont {A.}~\bibnamefont {Banerjee}},
  \ and\ \bibinfo {author} {\bibfnamefont {A.}~\bibnamefont {Narayan}},\ }\href
  {\doibase 10.1103/PhysRevA.103.L051101} {\bibfield  {journal} {\bibinfo
  {journal} {Phys. Rev. A}\ }\textbf {\bibinfo {volume} {103}},\ \bibinfo
  {pages} {L051101} (\bibinfo {year} {2021})}\BibitemShut {NoStop}%
\bibitem [{\citenamefont {Yang}\ and\ \citenamefont
  {Hu}(2019)}]{PhysRevB.99.081102}%
  \BibitemOpen
  \bibfield  {author} {\bibinfo {author} {\bibfnamefont {Z.}~\bibnamefont
  {Yang}}\ and\ \bibinfo {author} {\bibfnamefont {J.}~\bibnamefont {Hu}},\
  }\href {\doibase 10.1103/PhysRevB.99.081102} {\bibfield  {journal} {\bibinfo
  {journal} {Phys. Rev. B}\ }\textbf {\bibinfo {volume} {99}},\ \bibinfo
  {pages} {081102(R)} (\bibinfo {year} {2019})}\BibitemShut {NoStop}%
\bibitem [{\citenamefont {Carlstr\"om}\ \emph {et~al.}(2019)\citenamefont
  {Carlstr\"om}, \citenamefont {St\aa{}lhammar}, \citenamefont {Budich},\ and\
  \citenamefont {Bergholtz}}]{PhysRevB.99.161115}%
  \BibitemOpen
  \bibfield  {author} {\bibinfo {author} {\bibfnamefont {J.}~\bibnamefont
  {Carlstr\"om}}, \bibinfo {author} {\bibfnamefont {M.}~\bibnamefont
  {St\aa{}lhammar}}, \bibinfo {author} {\bibfnamefont {J.~C.}\ \bibnamefont
  {Budich}}, \ and\ \bibinfo {author} {\bibfnamefont {E.~J.}\ \bibnamefont
  {Bergholtz}},\ }\href {\doibase 10.1103/PhysRevB.99.161115} {\bibfield
  {journal} {\bibinfo  {journal} {Phys. Rev. B}\ }\textbf {\bibinfo {volume}
  {99}},\ \bibinfo {pages} {161115(R)} (\bibinfo {year} {2019})}\BibitemShut
  {NoStop}%
\bibitem [{\citenamefont {Yang}\ \emph
  {et~al.}(2020{\natexlab{b}})\citenamefont {Yang}, \citenamefont {Chiu},
  \citenamefont {Fang},\ and\ \citenamefont {Hu}}]{PhysRevLett.124.186402}%
  \BibitemOpen
  \bibfield  {author} {\bibinfo {author} {\bibfnamefont {Z.}~\bibnamefont
  {Yang}}, \bibinfo {author} {\bibfnamefont {C.-K.}\ \bibnamefont {Chiu}},
  \bibinfo {author} {\bibfnamefont {C.}~\bibnamefont {Fang}}, \ and\ \bibinfo
  {author} {\bibfnamefont {J.}~\bibnamefont {Hu}},\ }\href {\doibase
  10.1103/PhysRevLett.124.186402} {\bibfield  {journal} {\bibinfo  {journal}
  {Phys. Rev. Lett.}\ }\textbf {\bibinfo {volume} {124}},\ \bibinfo {pages}
  {186402} (\bibinfo {year} {2020}{\natexlab{b}})}\BibitemShut {NoStop}%
\bibitem [{\citenamefont {Eckardt}(2017)}]{RevModPhys.89.011004}%
  \BibitemOpen
  \bibfield  {author} {\bibinfo {author} {\bibfnamefont {A.}~\bibnamefont
  {Eckardt}},\ }\href {\doibase 10.1103/RevModPhys.89.011004} {\bibfield
  {journal} {\bibinfo  {journal} {Rev. Mod. Phys.}\ }\textbf {\bibinfo {volume}
  {89}},\ \bibinfo {pages} {011004} (\bibinfo {year} {2017})}\BibitemShut
  {NoStop}%
\bibitem [{\citenamefont {Meinert}\ \emph {et~al.}(2016)\citenamefont
  {Meinert}, \citenamefont {Mark}, \citenamefont {Lauber}, \citenamefont
  {Daley},\ and\ \citenamefont {N\"agerl}}]{PhysRevLett.116.205301}%
  \BibitemOpen
  \bibfield  {author} {\bibinfo {author} {\bibfnamefont {F.}~\bibnamefont
  {Meinert}}, \bibinfo {author} {\bibfnamefont {M.~J.}\ \bibnamefont {Mark}},
  \bibinfo {author} {\bibfnamefont {K.}~\bibnamefont {Lauber}}, \bibinfo
  {author} {\bibfnamefont {A.~J.}\ \bibnamefont {Daley}}, \ and\ \bibinfo
  {author} {\bibfnamefont {H.-C.}\ \bibnamefont {N\"agerl}},\ }\href {\doibase
  10.1103/PhysRevLett.116.205301} {\bibfield  {journal} {\bibinfo  {journal}
  {Phys. Rev. Lett.}\ }\textbf {\bibinfo {volume} {116}},\ \bibinfo {pages}
  {205301} (\bibinfo {year} {2016})}\BibitemShut {NoStop}%
\bibitem [{\citenamefont {Rechtsman}\ \emph {et~al.}(2013)\citenamefont
  {Rechtsman}, \citenamefont {Zeuner}, \citenamefont {Plotnik}, \citenamefont
  {Lumer}, \citenamefont {Podolsky}, \citenamefont {Dreisow}, \citenamefont
  {Nolte}, \citenamefont {Segev},\ and\ \citenamefont
  {Szameit}}]{Rechtsman2013}%
  \BibitemOpen
  \bibfield  {author} {\bibinfo {author} {\bibfnamefont {M.~C.}\ \bibnamefont
  {Rechtsman}}, \bibinfo {author} {\bibfnamefont {J.~M.}\ \bibnamefont
  {Zeuner}}, \bibinfo {author} {\bibfnamefont {Y.}~\bibnamefont {Plotnik}},
  \bibinfo {author} {\bibfnamefont {Y.}~\bibnamefont {Lumer}}, \bibinfo
  {author} {\bibfnamefont {D.}~\bibnamefont {Podolsky}}, \bibinfo {author}
  {\bibfnamefont {F.}~\bibnamefont {Dreisow}}, \bibinfo {author} {\bibfnamefont
  {S.}~\bibnamefont {Nolte}}, \bibinfo {author} {\bibfnamefont
  {M.}~\bibnamefont {Segev}}, \ and\ \bibinfo {author} {\bibfnamefont
  {A.}~\bibnamefont {Szameit}},\ }\href {\doibase 10.1038/nature12066}
  {\bibfield  {journal} {\bibinfo  {journal} {Nature (London)}\ }\textbf
  {\bibinfo {volume} {496}},\ \bibinfo {pages} {196} (\bibinfo {year}
  {2013})}\BibitemShut {NoStop}%
\bibitem [{\citenamefont {Cheng}\ \emph {et~al.}(2019)\citenamefont {Cheng},
  \citenamefont {Pan}, \citenamefont {Wang}, \citenamefont {Zhang},
  \citenamefont {Yu}, \citenamefont {Gover}, \citenamefont {Zhang},
  \citenamefont {Li}, \citenamefont {Zhou},\ and\ \citenamefont
  {Zhu}}]{PhysRevLett.122.173901}%
  \BibitemOpen
  \bibfield  {author} {\bibinfo {author} {\bibfnamefont {Q.}~\bibnamefont
  {Cheng}}, \bibinfo {author} {\bibfnamefont {Y.}~\bibnamefont {Pan}}, \bibinfo
  {author} {\bibfnamefont {H.}~\bibnamefont {Wang}}, \bibinfo {author}
  {\bibfnamefont {C.}~\bibnamefont {Zhang}}, \bibinfo {author} {\bibfnamefont
  {D.}~\bibnamefont {Yu}}, \bibinfo {author} {\bibfnamefont {A.}~\bibnamefont
  {Gover}}, \bibinfo {author} {\bibfnamefont {H.}~\bibnamefont {Zhang}},
  \bibinfo {author} {\bibfnamefont {T.}~\bibnamefont {Li}}, \bibinfo {author}
  {\bibfnamefont {L.}~\bibnamefont {Zhou}}, \ and\ \bibinfo {author}
  {\bibfnamefont {S.}~\bibnamefont {Zhu}},\ }\href {\doibase
  10.1103/PhysRevLett.122.173901} {\bibfield  {journal} {\bibinfo  {journal}
  {Phys. Rev. Lett.}\ }\textbf {\bibinfo {volume} {122}},\ \bibinfo {pages}
  {173901} (\bibinfo {year} {2019})}\BibitemShut {NoStop}%
\bibitem [{\citenamefont {Roushan}\ \emph {et~al.}(2017)\citenamefont
  {Roushan}, \citenamefont {Neill}, \citenamefont {Megrant}, \citenamefont
  {Chen}, \citenamefont {Babbush}, \citenamefont {Barends}, \citenamefont
  {Campbell}, \citenamefont {Chen}, \citenamefont {Chiaro}, \citenamefont
  {Dunsworth}, \citenamefont {Fowler}, \citenamefont {Jeffrey}, \citenamefont
  {Kelly}, \citenamefont {Lucero}, \citenamefont {Mutus}, \citenamefont {OHuo
  Heng~alley}, \citenamefont {Neeley}, \citenamefont {Quintana}, \citenamefont
  {Sank}, \citenamefont {Vainsencher}, \citenamefont {Wenner}, \citenamefont
  {White}, \citenamefont {Kapit}, \citenamefont {Neven},\ and\ \citenamefont
  {Martinis}}]{Roushan2017}%
  \BibitemOpen
  \bibfield  {author} {\bibinfo {author} {\bibfnamefont {P.}~\bibnamefont
  {Roushan}}, \bibinfo {author} {\bibfnamefont {C.}~\bibnamefont {Neill}},
  \bibinfo {author} {\bibfnamefont {A.}~\bibnamefont {Megrant}}, \bibinfo
  {author} {\bibfnamefont {Y.}~\bibnamefont {Chen}}, \bibinfo {author}
  {\bibfnamefont {R.}~\bibnamefont {Babbush}}, \bibinfo {author} {\bibfnamefont
  {R.}~\bibnamefont {Barends}}, \bibinfo {author} {\bibfnamefont
  {B.}~\bibnamefont {Campbell}}, \bibinfo {author} {\bibfnamefont
  {Z.}~\bibnamefont {Chen}}, \bibinfo {author} {\bibfnamefont {B.}~\bibnamefont
  {Chiaro}}, \bibinfo {author} {\bibfnamefont {A.}~\bibnamefont {Dunsworth}},
  \bibinfo {author} {\bibfnamefont {A.}~\bibnamefont {Fowler}}, \bibinfo
  {author} {\bibfnamefont {E.}~\bibnamefont {Jeffrey}}, \bibinfo {author}
  {\bibfnamefont {J.}~\bibnamefont {Kelly}}, \bibinfo {author} {\bibfnamefont
  {E.}~\bibnamefont {Lucero}}, \bibinfo {author} {\bibfnamefont
  {J.}~\bibnamefont {Mutus}}, \bibinfo {author} {\bibfnamefont {P.~J.~J.}\
  \bibnamefont {OHuo Heng~alley}}, \bibinfo {author} {\bibfnamefont
  {M.}~\bibnamefont {Neeley}}, \bibinfo {author} {\bibfnamefont
  {C.}~\bibnamefont {Quintana}}, \bibinfo {author} {\bibfnamefont
  {D.}~\bibnamefont {Sank}}, \bibinfo {author} {\bibfnamefont {A.}~\bibnamefont
  {Vainsencher}}, \bibinfo {author} {\bibfnamefont {J.}~\bibnamefont {Wenner}},
  \bibinfo {author} {\bibfnamefont {T.}~\bibnamefont {White}}, \bibinfo
  {author} {\bibfnamefont {E.}~\bibnamefont {Kapit}}, \bibinfo {author}
  {\bibfnamefont {H.}~\bibnamefont {Neven}}, \ and\ \bibinfo {author}
  {\bibfnamefont {J.}~\bibnamefont {Martinis}},\ }\href {\doibase
  10.1038/nphys3930} {\bibfield  {journal} {\bibinfo  {journal} {Nature
  Physics}\ }\textbf {\bibinfo {volume} {13}},\ \bibinfo {pages} {146}
  (\bibinfo {year} {2017})}\BibitemShut {NoStop}%
\bibitem [{\citenamefont {Chitsazi}\ \emph {et~al.}(2017)\citenamefont
  {Chitsazi}, \citenamefont {Li}, \citenamefont {Ellis},\ and\ \citenamefont
  {Kottos}}]{PhysRevLett.119.093901}%
  \BibitemOpen
  \bibfield  {author} {\bibinfo {author} {\bibfnamefont {M.}~\bibnamefont
  {Chitsazi}}, \bibinfo {author} {\bibfnamefont {H.}~\bibnamefont {Li}},
  \bibinfo {author} {\bibfnamefont {F.~M.}\ \bibnamefont {Ellis}}, \ and\
  \bibinfo {author} {\bibfnamefont {T.}~\bibnamefont {Kottos}},\ }\href
  {\doibase 10.1103/PhysRevLett.119.093901} {\bibfield  {journal} {\bibinfo
  {journal} {Phys. Rev. Lett.}\ }\textbf {\bibinfo {volume} {119}},\ \bibinfo
  {pages} {093901} (\bibinfo {year} {2017})}\BibitemShut {NoStop}%
\bibitem [{\citenamefont {McIver}\ \emph {et~al.}(2020)\citenamefont {McIver},
  \citenamefont {Schulte}, \citenamefont {Stein}, \citenamefont {Matsuyama},
  \citenamefont {Jotzu}, \citenamefont {Meier},\ and\ \citenamefont
  {Cavalleri}}]{McIver2020}%
  \BibitemOpen
  \bibfield  {author} {\bibinfo {author} {\bibfnamefont {J.~W.}\ \bibnamefont
  {McIver}}, \bibinfo {author} {\bibfnamefont {B.}~\bibnamefont {Schulte}},
  \bibinfo {author} {\bibfnamefont {F.-U.}\ \bibnamefont {Stein}}, \bibinfo
  {author} {\bibfnamefont {T.}~\bibnamefont {Matsuyama}}, \bibinfo {author}
  {\bibfnamefont {G.}~\bibnamefont {Jotzu}}, \bibinfo {author} {\bibfnamefont
  {G.}~\bibnamefont {Meier}}, \ and\ \bibinfo {author} {\bibfnamefont
  {A.}~\bibnamefont {Cavalleri}},\ }\href {\doibase 10.1038/s41567-019-0698-y}
  {\bibfield  {journal} {\bibinfo  {journal} {Nature Physics}\ }\textbf
  {\bibinfo {volume} {16}},\ \bibinfo {pages} {38} (\bibinfo {year}
  {2020})}\BibitemShut {NoStop}%
\bibitem [{\citenamefont {Yan}\ and\ \citenamefont
  {Wang}(2016)}]{PhysRevLett.117.087402}%
  \BibitemOpen
  \bibfield  {author} {\bibinfo {author} {\bibfnamefont {Z.}~\bibnamefont
  {Yan}}\ and\ \bibinfo {author} {\bibfnamefont {Z.}~\bibnamefont {Wang}},\
  }\href {\doibase 10.1103/PhysRevLett.117.087402} {\bibfield  {journal}
  {\bibinfo  {journal} {Phys. Rev. Lett.}\ }\textbf {\bibinfo {volume} {117}},\
  \bibinfo {pages} {087402} (\bibinfo {year} {2016})}\BibitemShut {NoStop}%
\bibitem [{\citenamefont {He}\ and\ \citenamefont
  {Huang}(2020)}]{PhysRevA.102.062201}%
  \BibitemOpen
  \bibfield  {author} {\bibinfo {author} {\bibfnamefont {P.}~\bibnamefont
  {He}}\ and\ \bibinfo {author} {\bibfnamefont {Z.-H.}\ \bibnamefont {Huang}},\
  }\href {\doibase 10.1103/PhysRevA.102.062201} {\bibfield  {journal} {\bibinfo
   {journal} {Phys. Rev. A}\ }\textbf {\bibinfo {volume} {102}},\ \bibinfo
  {pages} {062201} (\bibinfo {year} {2020})}\BibitemShut {NoStop}%
\bibitem [{\citenamefont {Wu}\ and\ \citenamefont
  {An}(2020)}]{PhysRevB.102.041119}%
  \BibitemOpen
  \bibfield  {author} {\bibinfo {author} {\bibfnamefont {H.}~\bibnamefont
  {Wu}}\ and\ \bibinfo {author} {\bibfnamefont {J.-H.}\ \bibnamefont {An}},\
  }\href {\doibase 10.1103/PhysRevB.102.041119} {\bibfield  {journal} {\bibinfo
   {journal} {Phys. Rev. B}\ }\textbf {\bibinfo {volume} {102}},\ \bibinfo
  {pages} {041119(R)} (\bibinfo {year} {2020})}\BibitemShut {NoStop}%
\bibitem [{\citenamefont {Zhou}\ and\ \citenamefont
  {Pan}(2019)}]{PhysRevA.100.053608}%
  \BibitemOpen
  \bibfield  {author} {\bibinfo {author} {\bibfnamefont {L.}~\bibnamefont
  {Zhou}}\ and\ \bibinfo {author} {\bibfnamefont {J.}~\bibnamefont {Pan}},\
  }\href {\doibase 10.1103/PhysRevA.100.053608} {\bibfield  {journal} {\bibinfo
   {journal} {Phys. Rev. A}\ }\textbf {\bibinfo {volume} {100}},\ \bibinfo
  {pages} {053608} (\bibinfo {year} {2019})}\BibitemShut {NoStop}%
\bibitem [{\citenamefont {Li}\ \emph {et~al.}(2019)\citenamefont {Li},
  \citenamefont {Ni}, \citenamefont {Weiner}, \citenamefont {Al\`u},\ and\
  \citenamefont {Khanikaev}}]{PhysRevB.100.045423}%
  \BibitemOpen
  \bibfield  {author} {\bibinfo {author} {\bibfnamefont {M.}~\bibnamefont
  {Li}}, \bibinfo {author} {\bibfnamefont {X.}~\bibnamefont {Ni}}, \bibinfo
  {author} {\bibfnamefont {M.}~\bibnamefont {Weiner}}, \bibinfo {author}
  {\bibfnamefont {A.}~\bibnamefont {Al\`u}}, \ and\ \bibinfo {author}
  {\bibfnamefont {A.~B.}\ \bibnamefont {Khanikaev}},\ }\href {\doibase
  10.1103/PhysRevB.100.045423} {\bibfield  {journal} {\bibinfo  {journal}
  {Phys. Rev. B}\ }\textbf {\bibinfo {volume} {100}},\ \bibinfo {pages}
  {045423} (\bibinfo {year} {2019})}\BibitemShut {NoStop}%
\bibitem [{\citenamefont {H\"ockendorf}\ \emph {et~al.}(2019)\citenamefont
  {H\"ockendorf}, \citenamefont {Alvermann},\ and\ \citenamefont
  {Fehske}}]{PhysRevLett.123.190403}%
  \BibitemOpen
  \bibfield  {author} {\bibinfo {author} {\bibfnamefont {B.}~\bibnamefont
  {H\"ockendorf}}, \bibinfo {author} {\bibfnamefont {A.}~\bibnamefont
  {Alvermann}}, \ and\ \bibinfo {author} {\bibfnamefont {H.}~\bibnamefont
  {Fehske}},\ }\href {\doibase 10.1103/PhysRevLett.123.190403} {\bibfield
  {journal} {\bibinfo  {journal} {Phys. Rev. Lett.}\ }\textbf {\bibinfo
  {volume} {123}},\ \bibinfo {pages} {190403} (\bibinfo {year}
  {2019})}\BibitemShut {NoStop}%
\bibitem [{\citenamefont {Zhang}\ and\ \citenamefont
  {Gong}(2020)}]{PhysRevB.101.045415}%
  \BibitemOpen
  \bibfield  {author} {\bibinfo {author} {\bibfnamefont {X.}~\bibnamefont
  {Zhang}}\ and\ \bibinfo {author} {\bibfnamefont {J.}~\bibnamefont {Gong}},\
  }\href {\doibase 10.1103/PhysRevB.101.045415} {\bibfield  {journal} {\bibinfo
   {journal} {Phys. Rev. B}\ }\textbf {\bibinfo {volume} {101}},\ \bibinfo
  {pages} {045415} (\bibinfo {year} {2020})}\BibitemShut {NoStop}%
\bibitem [{\citenamefont {Wu}\ \emph {et~al.}(2021)\citenamefont {Wu},
  \citenamefont {Wang},\ and\ \citenamefont {An}}]{PhysRevB.103.L041115}%
  \BibitemOpen
  \bibfield  {author} {\bibinfo {author} {\bibfnamefont {H.}~\bibnamefont
  {Wu}}, \bibinfo {author} {\bibfnamefont {B.-Q.}\ \bibnamefont {Wang}}, \ and\
  \bibinfo {author} {\bibfnamefont {J.-H.}\ \bibnamefont {An}},\ }\href
  {\doibase 10.1103/PhysRevB.103.L041115} {\bibfield  {journal} {\bibinfo
  {journal} {Phys. Rev. B}\ }\textbf {\bibinfo {volume} {103}},\ \bibinfo
  {pages} {L041115} (\bibinfo {year} {2021})}\BibitemShut {NoStop}%
\bibitem [{\citenamefont {Bucciantini}\ \emph {et~al.}(2017)\citenamefont
  {Bucciantini}, \citenamefont {Roy}, \citenamefont {Kitamura},\ and\
  \citenamefont {Oka}}]{PhysRevB.96.041126}%
  \BibitemOpen
  \bibfield  {author} {\bibinfo {author} {\bibfnamefont {L.}~\bibnamefont
  {Bucciantini}}, \bibinfo {author} {\bibfnamefont {S.}~\bibnamefont {Roy}},
  \bibinfo {author} {\bibfnamefont {S.}~\bibnamefont {Kitamura}}, \ and\
  \bibinfo {author} {\bibfnamefont {T.}~\bibnamefont {Oka}},\ }\href {\doibase
  10.1103/PhysRevB.96.041126} {\bibfield  {journal} {\bibinfo  {journal} {Phys.
  Rev. B}\ }\textbf {\bibinfo {volume} {96}},\ \bibinfo {pages} {041126(R)}
  (\bibinfo {year} {2017})}\BibitemShut {NoStop}%
\bibitem [{\citenamefont {Tong}\ \emph {et~al.}(2013)\citenamefont {Tong},
  \citenamefont {An}, \citenamefont {Gong}, \citenamefont {Luo},\ and\
  \citenamefont {Oh}}]{PhysRevB.87.201109}%
  \BibitemOpen
  \bibfield  {author} {\bibinfo {author} {\bibfnamefont {Q.-J.}\ \bibnamefont
  {Tong}}, \bibinfo {author} {\bibfnamefont {J.-H.}\ \bibnamefont {An}},
  \bibinfo {author} {\bibfnamefont {J.}~\bibnamefont {Gong}}, \bibinfo {author}
  {\bibfnamefont {H.-G.}\ \bibnamefont {Luo}}, \ and\ \bibinfo {author}
  {\bibfnamefont {C.~H.}\ \bibnamefont {Oh}},\ }\href {\doibase
  10.1103/PhysRevB.87.201109} {\bibfield  {journal} {\bibinfo  {journal} {Phys.
  Rev. B}\ }\textbf {\bibinfo {volume} {87}},\ \bibinfo {pages} {201109(R)}
  (\bibinfo {year} {2013})}\BibitemShut {NoStop}%
\bibitem [{\citenamefont {Xiong}\ \emph {et~al.}(2016)\citenamefont {Xiong},
  \citenamefont {Gong},\ and\ \citenamefont {An}}]{PhysRevB.93.184306}%
  \BibitemOpen
  \bibfield  {author} {\bibinfo {author} {\bibfnamefont {T.-S.}\ \bibnamefont
  {Xiong}}, \bibinfo {author} {\bibfnamefont {J.}~\bibnamefont {Gong}}, \ and\
  \bibinfo {author} {\bibfnamefont {J.-H.}\ \bibnamefont {An}},\ }\href
  {\doibase 10.1103/PhysRevB.93.184306} {\bibfield  {journal} {\bibinfo
  {journal} {Phys. Rev. B}\ }\textbf {\bibinfo {volume} {93}},\ \bibinfo
  {pages} {184306} (\bibinfo {year} {2016})}\BibitemShut {NoStop}%
\bibitem [{\citenamefont {Liu}\ \emph {et~al.}(2019)\citenamefont {Liu},
  \citenamefont {Xiong}, \citenamefont {Zhang},\ and\ \citenamefont
  {An}}]{PhysRevA.100.023622}%
  \BibitemOpen
  \bibfield  {author} {\bibinfo {author} {\bibfnamefont {H.}~\bibnamefont
  {Liu}}, \bibinfo {author} {\bibfnamefont {T.-S.}\ \bibnamefont {Xiong}},
  \bibinfo {author} {\bibfnamefont {W.}~\bibnamefont {Zhang}}, \ and\ \bibinfo
  {author} {\bibfnamefont {J.-H.}\ \bibnamefont {An}},\ }\href {\doibase
  10.1103/PhysRevA.100.023622} {\bibfield  {journal} {\bibinfo  {journal}
  {Phys. Rev. A}\ }\textbf {\bibinfo {volume} {100}},\ \bibinfo {pages}
  {023622} (\bibinfo {year} {2019})}\BibitemShut {NoStop}%
\bibitem [{\citenamefont {Li}\ \emph {et~al.}(2018{\natexlab{b}})\citenamefont
  {Li}, \citenamefont {Lee},\ and\ \citenamefont
  {Gong}}]{PhysRevLett.121.036401}%
  \BibitemOpen
  \bibfield  {author} {\bibinfo {author} {\bibfnamefont {L.}~\bibnamefont
  {Li}}, \bibinfo {author} {\bibfnamefont {C.~H.}\ \bibnamefont {Lee}}, \ and\
  \bibinfo {author} {\bibfnamefont {J.}~\bibnamefont {Gong}},\ }\href {\doibase
  10.1103/PhysRevLett.121.036401} {\bibfield  {journal} {\bibinfo  {journal}
  {Phys. Rev. Lett.}\ }\textbf {\bibinfo {volume} {121}},\ \bibinfo {pages}
  {036401} (\bibinfo {year} {2018}{\natexlab{b}})}\BibitemShut {NoStop}%
\bibitem [{\citenamefont {Banerjee}\ and\ \citenamefont
  {Narayan}(2020)}]{PhysRevB.102.205423}%
  \BibitemOpen
  \bibfield  {author} {\bibinfo {author} {\bibfnamefont {A.}~\bibnamefont
  {Banerjee}}\ and\ \bibinfo {author} {\bibfnamefont {A.}~\bibnamefont
  {Narayan}},\ }\href {\doibase 10.1103/PhysRevB.102.205423} {\bibfield
  {journal} {\bibinfo  {journal} {Phys. Rev. B}\ }\textbf {\bibinfo {volume}
  {102}},\ \bibinfo {pages} {205423} (\bibinfo {year} {2020})}\BibitemShut
  {NoStop}%
\bibitem [{\citenamefont {Zhang}\ \emph {et~al.}(2015)\citenamefont {Zhang},
  \citenamefont {Zhu},\ and\ \citenamefont {Wang}}]{PhysRevA.92.013632}%
  \BibitemOpen
  \bibfield  {author} {\bibinfo {author} {\bibfnamefont {D.-W.}\ \bibnamefont
  {Zhang}}, \bibinfo {author} {\bibfnamefont {S.-L.}\ \bibnamefont {Zhu}}, \
  and\ \bibinfo {author} {\bibfnamefont {Z.~D.}\ \bibnamefont {Wang}},\ }\href
  {\doibase 10.1103/PhysRevA.92.013632} {\bibfield  {journal} {\bibinfo
  {journal} {Phys. Rev. A}\ }\textbf {\bibinfo {volume} {92}},\ \bibinfo
  {pages} {013632} (\bibinfo {year} {2015})}\BibitemShut {NoStop}%
\bibitem [{sup()}]{supplementary}%
  \BibitemOpen
  \href@noop {} {}See the Supplemental Material for a detailed derivation of
  the generalized Brillouin zone for both of the static and the periodically
  driven cases.\BibitemShut {Stop}%
\bibitem [{\citenamefont {Hirsbrunner}\ \emph {et~al.}(2019)\citenamefont
  {Hirsbrunner}, \citenamefont {Philip},\ and\ \citenamefont
  {Gilbert}}]{PhysRevB.100.081104}%
  \BibitemOpen
  \bibfield  {author} {\bibinfo {author} {\bibfnamefont {M.~R.}\ \bibnamefont
  {Hirsbrunner}}, \bibinfo {author} {\bibfnamefont {T.~M.}\ \bibnamefont
  {Philip}}, \ and\ \bibinfo {author} {\bibfnamefont {M.~J.}\ \bibnamefont
  {Gilbert}},\ }\href {\doibase 10.1103/PhysRevB.100.081104} {\bibfield
  {journal} {\bibinfo  {journal} {Phys. Rev. B}\ }\textbf {\bibinfo {volume}
  {100}},\ \bibinfo {pages} {081104(R)} (\bibinfo {year} {2019})}\BibitemShut
  {NoStop}%
\bibitem [{\citenamefont {Choi}\ \emph {et~al.}(2017)\citenamefont {Choi},
  \citenamefont {Choi}, \citenamefont {Landig}, \citenamefont {Kucsko},
  \citenamefont {Zhou}, \citenamefont {Isoya}, \citenamefont {Jelezko},
  \citenamefont {Onoda}, \citenamefont {Sumiya}, \citenamefont {Khemani},
  \citenamefont {von Keyserlingk}, \citenamefont {Yao}, \citenamefont
  {Demler},\ and\ \citenamefont {Lukin}}]{Choi2017}%
  \BibitemOpen
  \bibfield  {author} {\bibinfo {author} {\bibfnamefont {S.}~\bibnamefont
  {Choi}}, \bibinfo {author} {\bibfnamefont {J.}~\bibnamefont {Choi}}, \bibinfo
  {author} {\bibfnamefont {R.}~\bibnamefont {Landig}}, \bibinfo {author}
  {\bibfnamefont {G.}~\bibnamefont {Kucsko}}, \bibinfo {author} {\bibfnamefont
  {H.}~\bibnamefont {Zhou}}, \bibinfo {author} {\bibfnamefont {J.}~\bibnamefont
  {Isoya}}, \bibinfo {author} {\bibfnamefont {F.}~\bibnamefont {Jelezko}},
  \bibinfo {author} {\bibfnamefont {S.}~\bibnamefont {Onoda}}, \bibinfo
  {author} {\bibfnamefont {H.}~\bibnamefont {Sumiya}}, \bibinfo {author}
  {\bibfnamefont {V.}~\bibnamefont {Khemani}}, \bibinfo {author} {\bibfnamefont
  {C.}~\bibnamefont {von Keyserlingk}}, \bibinfo {author} {\bibfnamefont
  {N.~Y.}\ \bibnamefont {Yao}}, \bibinfo {author} {\bibfnamefont
  {E.}~\bibnamefont {Demler}}, \ and\ \bibinfo {author} {\bibfnamefont {M.~D.}\
  \bibnamefont {Lukin}},\ }\href {\doibase 10.1038/nature21426} {\bibfield
  {journal} {\bibinfo  {journal} {Nature (London)}\ }\textbf {\bibinfo {volume}
  {543}},\ \bibinfo {pages} {221} (\bibinfo {year} {2017})}\BibitemShut
  {NoStop}%
\bibitem [{\citenamefont {Zhang}\ \emph {et~al.}(2017)\citenamefont {Zhang},
  \citenamefont {Hess}, \citenamefont {Kyprianidis}, \citenamefont {Becker},
  \citenamefont {Lee}, \citenamefont {Smith}, \citenamefont {Pagano},
  \citenamefont {Potirniche}, \citenamefont {Potter}, \citenamefont
  {Vishwanath}, \citenamefont {Yao},\ and\ \citenamefont {Monroe}}]{Zhang2017}%
  \BibitemOpen
  \bibfield  {author} {\bibinfo {author} {\bibfnamefont {J.}~\bibnamefont
  {Zhang}}, \bibinfo {author} {\bibfnamefont {P.~W.}\ \bibnamefont {Hess}},
  \bibinfo {author} {\bibfnamefont {A.}~\bibnamefont {Kyprianidis}}, \bibinfo
  {author} {\bibfnamefont {P.}~\bibnamefont {Becker}}, \bibinfo {author}
  {\bibfnamefont {A.}~\bibnamefont {Lee}}, \bibinfo {author} {\bibfnamefont
  {J.}~\bibnamefont {Smith}}, \bibinfo {author} {\bibfnamefont
  {G.}~\bibnamefont {Pagano}}, \bibinfo {author} {\bibfnamefont {I.-D.}\
  \bibnamefont {Potirniche}}, \bibinfo {author} {\bibfnamefont {A.~C.}\
  \bibnamefont {Potter}}, \bibinfo {author} {\bibfnamefont {A.}~\bibnamefont
  {Vishwanath}}, \bibinfo {author} {\bibfnamefont {N.~Y.}\ \bibnamefont {Yao}},
  \ and\ \bibinfo {author} {\bibfnamefont {C.}~\bibnamefont {Monroe}},\ }\href
  {\doibase 10.1038/nature21413} {\bibfield  {journal} {\bibinfo  {journal}
  {Nature (London)}\ }\textbf {\bibinfo {volume} {543}},\ \bibinfo {pages}
  {217} (\bibinfo {year} {2017})}\BibitemShut {NoStop}%
\bibitem [{\citenamefont {Rudner}\ \emph {et~al.}(2013)\citenamefont {Rudner},
  \citenamefont {Lindner}, \citenamefont {Berg},\ and\ \citenamefont
  {Levin}}]{PhysRevX.3.031005}%
  \BibitemOpen
  \bibfield  {author} {\bibinfo {author} {\bibfnamefont {M.~S.}\ \bibnamefont
  {Rudner}}, \bibinfo {author} {\bibfnamefont {N.~H.}\ \bibnamefont {Lindner}},
  \bibinfo {author} {\bibfnamefont {E.}~\bibnamefont {Berg}}, \ and\ \bibinfo
  {author} {\bibfnamefont {M.}~\bibnamefont {Levin}},\ }\href {\doibase
  10.1103/PhysRevX.3.031005} {\bibfield  {journal} {\bibinfo  {journal} {Phys.
  Rev. X}\ }\textbf {\bibinfo {volume} {3}},\ \bibinfo {pages} {031005}
  (\bibinfo {year} {2013})}\BibitemShut {NoStop}%
\bibitem [{\citenamefont {Xu}\ \emph {et~al.}(2017)\citenamefont {Xu},
  \citenamefont {Wang},\ and\ \citenamefont {Duan}}]{PhysRevLett.118.045701}%
  \BibitemOpen
  \bibfield  {author} {\bibinfo {author} {\bibfnamefont {Y.}~\bibnamefont
  {Xu}}, \bibinfo {author} {\bibfnamefont {S.-T.}\ \bibnamefont {Wang}}, \ and\
  \bibinfo {author} {\bibfnamefont {L.-M.}\ \bibnamefont {Duan}},\ }\href
  {\doibase 10.1103/PhysRevLett.118.045701} {\bibfield  {journal} {\bibinfo
  {journal} {Phys. Rev. Lett.}\ }\textbf {\bibinfo {volume} {118}},\ \bibinfo
  {pages} {045701} (\bibinfo {year} {2017})}\BibitemShut {NoStop}%
\bibitem [{\citenamefont {Zhu}\ \emph {et~al.}(2019)\citenamefont {Zhu},
  \citenamefont {Yu}, \citenamefont {Wu}, \citenamefont {Zhang}, \citenamefont
  {Zhang}, \citenamefont {Zhang},\ and\ \citenamefont
  {Yang}}]{PhysRevB.100.161401}%
  \BibitemOpen
  \bibfield  {author} {\bibinfo {author} {\bibfnamefont {Z.}~\bibnamefont
  {Zhu}}, \bibinfo {author} {\bibfnamefont {Z.-M.}\ \bibnamefont {Yu}},
  \bibinfo {author} {\bibfnamefont {W.}~\bibnamefont {Wu}}, \bibinfo {author}
  {\bibfnamefont {L.}~\bibnamefont {Zhang}}, \bibinfo {author} {\bibfnamefont
  {W.}~\bibnamefont {Zhang}}, \bibinfo {author} {\bibfnamefont
  {F.}~\bibnamefont {Zhang}}, \ and\ \bibinfo {author} {\bibfnamefont {S.~A.}\
  \bibnamefont {Yang}},\ }\href {\doibase 10.1103/PhysRevB.100.161401}
  {\bibfield  {journal} {\bibinfo  {journal} {Phys. Rev. B}\ }\textbf {\bibinfo
  {volume} {100}},\ \bibinfo {pages} {161401(R)} (\bibinfo {year}
  {2019})}\BibitemShut {NoStop}%
\end{thebibliography}%


\begin{thebibliography}{3}%
\makeatletter
\providecommand \@ifxundefined [1]{%
 \@ifx{#1\undefined}
}%
\providecommand \@ifnum [1]{%
 \ifnum #1\expandafter \@firstoftwo
 \else \expandafter \@secondoftwo
 \fi
}%
\providecommand \@ifx [1]{%
 \ifx #1\expandafter \@firstoftwo
 \else \expandafter \@secondoftwo
 \fi
}%
\providecommand \natexlab [1]{#1}%
\providecommand \enquote  [1]{``#1''}%
\providecommand \bibnamefont  [1]{#1}%
\providecommand \bibfnamefont [1]{#1}%
\providecommand \citenamefont [1]{#1}%
\providecommand \href@noop [0]{\@secondoftwo}%
\providecommand \href [0]{\begingroup \@sanitize@url \@href}%
\providecommand \@href[1]{\@@startlink{#1}\@@href}%
\providecommand \@@href[1]{\endgroup#1\@@endlink}%
\providecommand \@sanitize@url [0]{\catcode `\\12\catcode `\$12\catcode
  `\&12\catcode `\#12\catcode `\^12\catcode `\_12\catcode `\%12\relax}%
\providecommand \@@startlink[1]{}%
\providecommand \@@endlink[0]{}%
\providecommand \url  [0]{\begingroup\@sanitize@url \@url }%
\providecommand \@url [1]{\endgroup\@href {#1}{\urlprefix }}%
\providecommand \urlprefix  [0]{URL }%
\providecommand \Eprint [0]{\href }%
\providecommand \doibase [0]{https://doi.org/}%
\providecommand \selectlanguage [0]{\@gobble}%
\providecommand \bibinfo  [0]{\@secondoftwo}%
\providecommand \bibfield  [0]{\@secondoftwo}%
\providecommand \translation [1]{[#1]}%
\providecommand \BibitemOpen [0]{}%
\providecommand \bibitemStop [0]{}%
\providecommand \bibitemNoStop [0]{.\EOS\space}%
\providecommand \EOS [0]{\spacefactor3000\relax}%
\providecommand \BibitemShut  [1]{\csname bibitem#1\endcsname}%
\let\auto@bib@innerbib\@empty
%</preamble>
\bibitem [{\citenamefont {Zhang}\ \emph {et~al.}(2015)\citenamefont {Zhang},
  \citenamefont {Zhu},\ and\ \citenamefont {Wang}}]{PhysRevA.92.013632}%
  \BibitemOpen
  \bibfield  {author} {\bibinfo {author} {\bibfnamefont {D.-W.}\ \bibnamefont
  {Zhang}}, \bibinfo {author} {\bibfnamefont {S.-L.}\ \bibnamefont {Zhu}},\
  and\ \bibinfo {author} {\bibfnamefont {Z.~D.}\ \bibnamefont {Wang}},\
  }\bibfield  {title} {\bibinfo {title} {Simulating and exploring {W}eyl
  semimetal physics with cold atoms in a two-dimensional optical lattice},\
  }\href {https://doi.org/10.1103/PhysRevA.92.013632} {\bibfield  {journal}
  {\bibinfo  {journal} {Phys. Rev. A}\ }\textbf {\bibinfo {volume} {92}},\
  \bibinfo {pages} {013632} (\bibinfo {year} {2015})}\BibitemShut {NoStop}%
\bibitem [{\citenamefont {Yao}\ and\ \citenamefont
  {Wang}(2018)}]{PhysRevLett.121.086803}%
  \BibitemOpen
  \bibfield  {author} {\bibinfo {author} {\bibfnamefont {S.}~\bibnamefont
  {Yao}}\ and\ \bibinfo {author} {\bibfnamefont {Z.}~\bibnamefont {Wang}},\
  }\bibfield  {title} {\bibinfo {title} {Edge states and topological invariants
  of non-{H}ermitian systems},\ }\href
  {https://doi.org/10.1103/PhysRevLett.121.086803} {\bibfield  {journal}
  {\bibinfo  {journal} {Phys. Rev. Lett.}\ }\textbf {\bibinfo {volume} {121}},\
  \bibinfo {pages} {086803} (\bibinfo {year} {2018})}\BibitemShut {NoStop}%
\bibitem [{\citenamefont {Yi}\ and\ \citenamefont
  {Yang}(2020)}]{PhysRevLett.125.186802}%
  \BibitemOpen
  \bibfield  {author} {\bibinfo {author} {\bibfnamefont {Y.}~\bibnamefont
  {Yi}}\ and\ \bibinfo {author} {\bibfnamefont {Z.}~\bibnamefont {Yang}},\
  }\bibfield  {title} {\bibinfo {title} {Non-hermitian skin modes induced by
  on-site dissipations and chiral tunneling effect},\ }\href
  {https://doi.org/10.1103/PhysRevLett.125.186802} {\bibfield  {journal}
  {\bibinfo  {journal} {Phys. Rev. Lett.}\ }\textbf {\bibinfo {volume} {125}},\
  \bibinfo {pages} {186802} (\bibinfo {year} {2020})}\BibitemShut {NoStop}%
\end{thebibliography}%
\end{document}

\title{Supplemental material for ``Non-Hermitian Weyl semimetal and its Floquet engineering''}
\author{Hong Wu\orcid{0000-0003-3276-7823}}
\affiliation{Lanzhou Center for Theoretical Physics, Key Laboratory of Theoretical Physics of Gansu Province, Lanzhou University, Lanzhou 730000, China}
\author{Jun-Hong An\orcid{0000-0002-3475-0729}}\email{anjhong@lzu.edu.cn}
\affiliation{Lanzhou Center for Theoretical Physics, Key Laboratory of Theoretical Physics of Gansu Province, Lanzhou University, Lanzhou 730000, China}

\maketitle

\section{Generalized Brillouin zone of the static system}
\begin{figure}[tbp]
\centering
\includegraphics[width=.85\columnwidth]{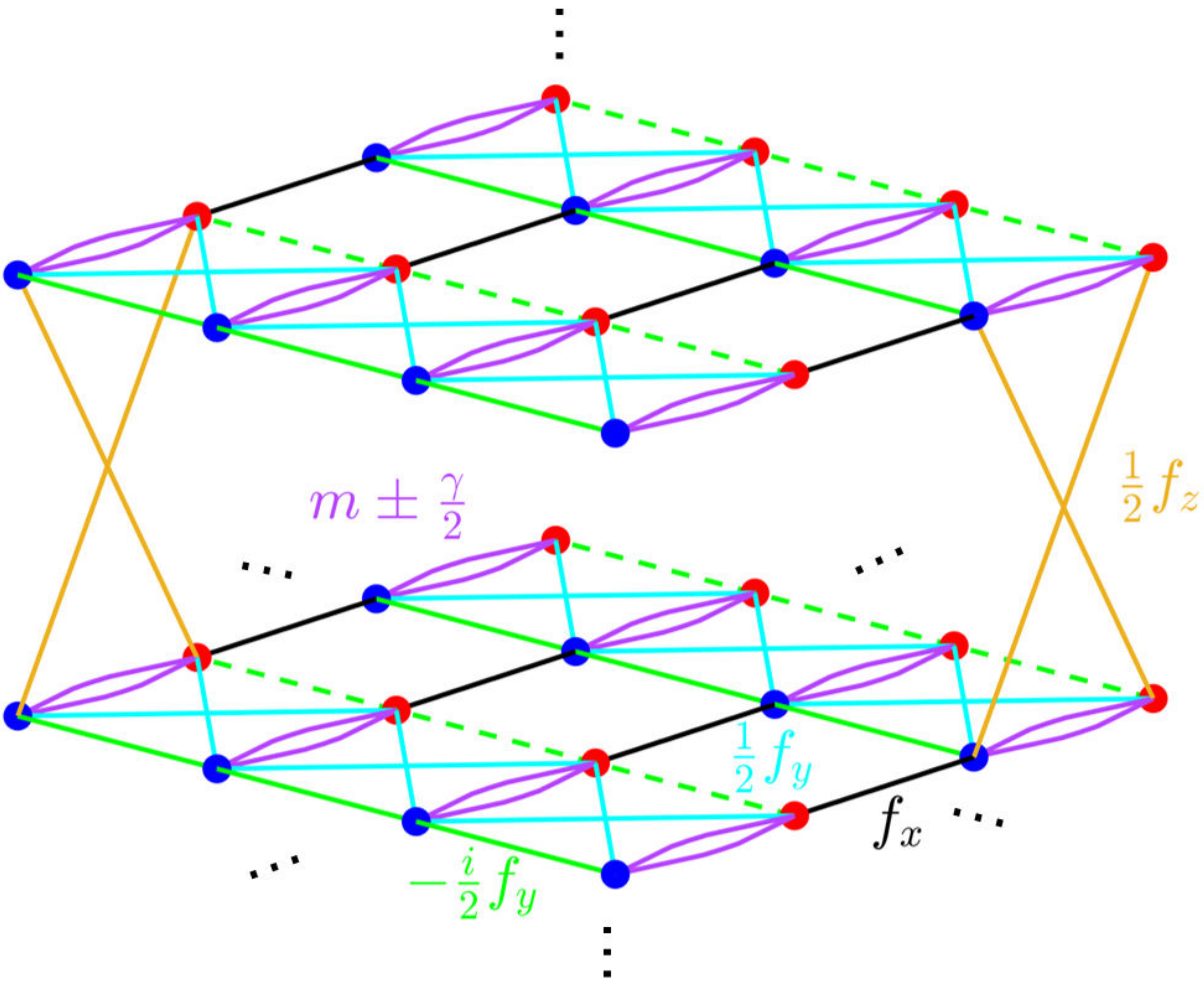}
\caption{Schematics of non-Hermitian Weyl semimetal on a cubic lattice with nonreciprocal intracell hopping rates $m\pm{\gamma/2}$, and intercell ones $f_x$ and $f_y$ in the single layer
and $f_z$ between two neighboring layers, respectively. The green dashed lines denote the hopping rates with a $\pi$-phase difference from the green solid lines.} \label{Model}
\end{figure}

We consider a non-Hermitian Weyl semimetal model on a cubic lattice as shown in Fig. \ref{Model}. Its Hamiltonian is
\begin{eqnarray}
\hat{H}&=&\sum_{\bf n}\Big\{{1\over 2}\Big[{f_x}\hat{C}^{\dag}_{{\bf n}+\bf{x}}(\tau_x-i{\tau_y})\hat{C}_{{\bf n}}+{f_y}\hat{C}^{\dag}_{{\bf n}+\bf{y}}(\tau_x-i{\tau_z})\hat{C}_{{\bf n}}\nonumber\\
&&+{f_z}\hat{C}^{\dag}_{{\bf n}+\bf{z}}\tau_x\hat{C}_{{\bf n}}+\text{H.c.}\Big]+\hat{C}^{\dag}_{{\bf n}}(m\tau_x+i\frac{\gamma}{2}\tau_y)\hat{C}_{{\bf n}}\Big\},\label{Haml}
\end{eqnarray}
where $\hat{C}_{\bf n}=(\hat{c}_{{\bf n},A},\hat{c}_{{\bf n},B})^T$, $\hat{c}_{{\bf n},j}$ is the annihilation operator at $j=A,B$ sublattice of the site ${\bf n}=(n_x,n_y,n_z)$, $f_\alpha$ are the hopping rates along $\alpha=x$, $y$, and $z$ directions with the unit vectors $\bf{x}$, $\bf{y}$, and $\bf{z}$, $\tau_\alpha$ are the Pauli matrices acting on the sublattice degrees of freedom, $m$ is the mass term, and the non-Hermitian term $\gamma$ is the nonreciprocal intracell hopping rate. Under the periodic-boundary condition, Eq. \eqref{Haml} reads $\hat{H}=\sum_{\bf k}{\hat{\bf C}}^\dag_{\bf k}\mathcal{H}({\bf k}){\hat{\bf C}}_{\bf k}$ with $\hat{\bf C}_{\bf k}=(\hat{\tilde c}_{{\bf k},A},\hat{\tilde{c}}_{{\bf k},B})^T$ and
\begin{eqnarray}
\mathcal{H}(\mathbf{k})&=&(f_x\cos k_x+F_{k_y,k_z})\sigma_x\nonumber\\
&&+(f_x\sin k_x+i\frac{\gamma}{2})\sigma_y+f_y\sin k_y\sigma_z,~~~~\label{Hamt}
\end{eqnarray}
where $\sigma_\alpha$ are the Pauli matrices and $F_{k_y,k_z}=m+f_y\cos k_y+f_z\cos k_z$. Related model in the Hermitian case was studied in \cite{PhysRevA.92.013632}.

The unique feature of the non-Hermitian system is the skin effect \cite{PhysRevLett.121.086803}, which breaks the bulk-boundary correspondence manifesting in the mismatching of the exceptional points under the open- and periodic-boundary conditions.
It can be readily seen from Fig. \ref{Model} that the non-Hermiticity introduced by the nonreciprocal intracell hopping only occurs along the $x$ direction. It naturally results in that the skin effect only exists in the $x$ direction. This conclusion can be confirmed by the symmetries of the system. Equation \eqref{Hamt} possesses the anomalous time-reversal symmetry along the $y$ direction, i.e.,
\begin{equation}
\sigma_x\mathcal{K}^T\mathcal{H}(k_x,k_y,k_z)\mathcal{K}^T\sigma_x=\mathcal{H}(k_x,-k_y,k_z),
\end{equation}
where $\mathcal{K}^T$ is the transpose operator, and the inversion symmetry alone the $z$ direction, i.e.,
\begin{equation}
\mathcal{H}(k_x,k_y,k_z)=\mathcal{H}(k_x,k_y,-k_z).
\end{equation}
It has been proven that the anomalous time-reversal symmetry or the inversion symmetry or both makes the skin effect absent \cite{PhysRevLett.125.186802}. Therefore, our system does not show the skin effect in the $y$ and $z$ directions.

The $y$ and $z$ directions can be well described by the conventional Brillouin zone, which can be obtained by the Fourier transform under the periodic boundary condition. Then Eq. \eqref{Haml} is converted into an effective one-dimensional non-Hermitian system parameterized by the conserved momenta $k_y$ and $k_z$, i.e.,
\begin{eqnarray}
\hat{H}&=&\sum_{k_y,k_z}\sum_{ n_x}[(F_{k_y,k_z}-\frac{\gamma}{2})\hat{b}^{\dag}_{n_x,k_y,k_z}\hat{a}_{n_x,k_y,k_z}\nonumber\\
&&+(F_{k_y,k_z}+\frac{\gamma}{2})\hat{a}^{\dag}_{n_x,k_y,k_z}\hat{b}_{n_x,k_y,k_z}\nonumber\\&&+f_y\sin k_y (\hat{a}_{n_x,k_y,k_z}^{\dag}\hat{a}_{n_x,k_y,k_z}-\hat{b}_{n_x,k_y,k_z}^{\dag}\hat{b}_{n_x,k_y,k_z})\nonumber\\
&&+f_x(\hat{a}^{\dag}_{n_x,k_y,k_z}\hat{b}_{n_x-1,k_y,k_z}+\text{H.c.})]. \label{static}
\end{eqnarray}
where $F_{k_y,k_z}=m+f_y\cos k_y+f_z\cos k_z$ and
\begin{eqnarray}
\hat{c}_{{\bf n},A}=\frac{1}{\sqrt{N_yN_z}}\sum_{k_y,k_z}e^{i(k_yn_y+k_zn_z)}\hat{a}_{n_x,k_y,k_z},\\
\hat{c}_{{\bf n},B}=\frac{1}{\sqrt{N_yN_z}}\sum_{k_y,k_z}e^{i(k_yn_y+k_zn_z)}\hat{b}_{n_x,k_y,k_z}.
\end{eqnarray} This is just the one-dimensional non-Hermitian Su-Schrieffer-Heeger model \cite{PhysRevLett.121.086803} with an additional $f_y$ term. According to the method developed in Ref. \cite{PhysRevLett.121.086803}, we can remove the skin effect by a similarity transformation
\begin{eqnarray}
\hat{\mathcal S}&=&\exp[\sum_{n_x}(\hat{a}_{n_x,k_y,k_z}^{\dag}\hat{a}_{n_x,k_y,k_z}\ln r^{n_x-1}\nonumber\\
&&+\hat{b}_{n_x,k_y,k_z}^{\dag}\hat{b}_{n_x,k_y,k_z}\ln r^{n_x})].
\end{eqnarray}
It converts Eq. \eqref{static} into $\hat{\bar{H}}=\hat{\mathcal S}^{-1}\hat{H}\hat{\mathcal S}$ with
\begin{eqnarray}
\hat{\bar{H}}&=&\sum_{k_y,k_z}\sum_{ n_x}[r^{-1}(F_{k_y,k_z}-\frac{\gamma}{2})\hat{b}^{\dag}_{n_x,k_y,k_z}\hat{a}_{n_x,k_y,k_z}\nonumber\\
&&+r(F_{k_y,k_z}+\frac{\gamma}{2})\hat{a}^{\dag}_{n_x,k_y,k_z}\hat{b}_{n_x,k_y,k_z}\nonumber\\&&+f_y\sin k_y (\hat{a}_{n_x,k_y,k_z}^{\dag}\hat{a}_{n_x,k_y,k_z}-\hat{b}_{n_x,k_y,k_z}^{\dag}\hat{b}_{n_x,k_y,k_z})\nonumber\\
&&+f_x(\hat{a}^{\dag}_{n_x,k_y,k_z}\hat{b}_{n_x-1,k_y,k_z}+\text{H.c.})].\label{hmtbar}
\end{eqnarray}
One can readily check that, as long as $r=|(F_{k_y,k_z}-\gamma/2)/(F_{k_y,k_z}+\gamma/2)|^{1/2}$, all the bulk states of $\hat{\bar H}$ do not reside in the edges anymore and thus the skin effect is removed \cite{PhysRevLett.121.086803}.
Rewriting Eqs. \eqref{hmtbar} in the momentum space, we have $\hat{\bar{H}}=\sum_{\bf k}{\bf C}_{\bf k}^\dag\hat{\bar{\mathcal H}}({\bf k}){\bf C}_{\bf k}$ with
\begin{eqnarray}
\hat{\bar{\mathcal H}}({\bf k})&=&
\left[\begin{array}{cc}
f_y\sin k_y & r(F_{k_y,k_z}+{\gamma\over 2}+{f_x\over\beta}) \\
r^{-1}(F_{k_y,k_z}-{\gamma\over 2}+f_x\beta)  & -f_y\sin k_y \\
\end{array}\right], \nonumber\\ \label{hamgbz}
\end{eqnarray}
where $\beta\equiv e^{i\tilde{k}_x}=re^{ik_x}$. Comparing Eq. \eqref{hamgbz} with Eq. (2) in the main text, we see that, except for the unimportant pre-factors $r^{\pm1}$, which do not affect the eigenvalues, they have the same form after replacing $k_x$ with $\tilde{k}_x$. This gives us an insight to remove the skin effect in the original non-Hermitian system by directly replacing ${k}_x$ in Eq. (2) of the main text by $\tilde{ k}_x$ \cite{PhysRevLett.121.086803}. Such complex vector $\tilde{ k}$ as well as $k_y$ and $k_z$ form a generalized Brillouin zone.

\section{Generalized Brillouin zone of the periodically driven system}
We can see that the similarity transformation $\hat{\mathcal S}$ to remove the skin effect is independent on $f_x$. When the periodic driving
\begin{equation}
f_x(t) =\begin{cases}q_1 f , ~t\in\lbrack zT, zT+T_1)\\q_2f, ~t\in\lbrack zT+T_1, (z+1)T)\end{cases}, \label{procotol}
\end{equation}
is applied on $f_x$, we readily have
\begin{equation}
\hat{\bar{U}}_{T}=\hat{\mathcal{S}}^{-1}\hat{U}_T\hat{\mathcal{S}}=e^{-i\hat{\bar{H}}_2T_2}e^{-i\hat{\bar{H}}_1T_1},
\end{equation}
where $\hat{\bar{H}}_j$ are Eq. \eqref{hmtbar} with $f_x=q_1 f$ and $q_2 f$, respectively. Since no skin effect is present in $\hat{\bar{H}}_j$, neither in $\hat{\bar{H}}_\text{eff}=\frac{i}{T}\ln[\hat{\bar{U}}_T]$. Therefore, we conclude that the generalized Brillouin zone of the periodically driven system is the same as that of static case.

\bibliography{ref}